\documentclass[prl,aps,superscriptaddress,twocolumn]{revtex4-2}

\usepackage{subfigure}
\usepackage{xcolor}
\usepackage{array}
\usepackage{booktabs}
\usepackage{graphicx}
\usepackage{dcolumn}
\usepackage{physics}
\usepackage{bm}
\usepackage{amsmath,amssymb,amsfonts,mathtools}
\usepackage{comment}
\newcommand{\ii}{\mathrm{i}}

\newcommand{\cL}{\mathcal{L}}

\newcommand{\normal}[1]{:\!#1\!:}

\usepackage[colorlinks=true]{hyperref}
\hypersetup{linkcolor=blue,citecolor=blue,urlcolor=blue}

\begin{document}

\title{Exact Lindbladian Dynamics from Conformal Embeddings and Topological Defects\\ in Conformal Field Theory}

\author{Chen Bai}
\email{baichen22@mails.ucas.ac.cn}
\affiliation{Kavli Institute for Theoretical Sciences, University of Chinese Academy of Sciences, Beijing 100190, China}

\date{\today}

\begin{abstract}
Analyzing the dynamics of physical observables in open quantum many-body systems is a fundamental but highly challenging task that has yielded very few exact results. In this work, we identify intrinsic conformal structures that restore exact solvability in $(1+1)$D conformal field theories. For \(N\) Majorana fermions with linear mode jumps, the adjoint Lindbladian is triangular on reduced even Majorana monomials, yielding recursive exact Heisenberg evolution. In Wess-Zumino-Witten models admitting conformal Majorana embeddings, this hierarchy gives exact dynamics of affine-current products realized as Majorana bilinears, including regimes where the Kac-Moody current algebra alone does not close. In diagonal rational conformal field theories, Verlinde topological defect lines furnish jump operators whose primary-sector dynamics is exactly diagonal: topological-charge probabilities are conserved, while intersector coherences dephase at rates fixed by the modular \(S\) matrix and nonnegative measurement strengths. These examples show that intrinsic conformal structures, such as conformal embeddings and modular data, can organize exactly solvable open conformal dynamics.
\end{abstract}

\maketitle

\textit{Introduction} Understanding open quantum many-body dynamics is a central problem in nonequilibrium physics, since environmental couplings govern decoherence, relaxation, transport, and dissipative state preparation. In Markovian regimes, these processes are described by Lindblad master equations in settings ranging from engineered atomic and optical platforms to quantum materials \cite{Muller2012AdvAMOP61,Reiter2017NatCommun8,Tomadin2012PRA86,BreuerPetruccione2007OpenQuantumSystems,Lindblad1976CMP48} and semiclassical field theory in de Sitter space \cite{Ishii2025PRD112}. Yet exact analytic control over interacting Lindbladians remains rare. Systematic solvability is largely confined to quasi-free fermionic or bosonic systems, where covariance-matrix and third-quantization methods reduce the dynamics to effectively single-particle data \cite{Prosen2008NJP10,ProsenZunkovic2010NJP12,BarthelZhang2022JSTAT2022,YamanakaSasamoto2023SciPost14}. For genuinely interacting systems, such reductions are unavailable, making exact solvability especially challenging. This motivates the search for broader classes of Lindbladians with analytically tractable dissipative dynamics.

Conformal field theory (CFT) provides a natural arena for this search: it describes critical quantum matter and often owes its solvability to enlarged conformal symmetry and operator-algebraic structures. Although non-unitary dynamics in CFTs has been studied extensively \cite{Lotkov2023PRB108,Mao2024JHEP2024,Su2024PRResearch6,Wen2024JSTAT2024,Lapierre2025PRB112,Lin2025PRB111,Bai2026Arxiv260109290,Barad2025PRB112,KarchWang2026Arxiv260116888,Lapierre:2025gyf,Lapierre:2025wya,toni2026effectivedynamicsopen2d}, Lindbladian dynamics in CFTs remains comparatively underdeveloped. A recent step in this direction was taken by Tang, Barad, and Wen, who introduced Wess-Zumino-Witten (WZW) current-mode Lindbladians with jump operators linear in Kac--Moody currents \cite{TangBaradWen2026}. They showed that affine current algebra yields closed Heisenberg operator dynamics broadly in Abelian U\((1)_k\) theories, and that in non-Abelian WZW theories the same closure survives in the single-current sector under symmetric-rate conditions. Their result establishes operator closure as an organizing principle for exactly solvable dissipative dynamics in CFT.

Motivated by this operator-closure viewpoint, we identify new families of exactly solvable Lindbladian dynamics in \((1+1)\)D CFTs. We show that linear Majorana-mode jumps generate a triangular hierarchy on reduced even Majorana monomials, giving recursive exact solutions and cooling dynamics. We then use conformal embeddings to transfer this hierarchy to WZW affine currents realized as Majorana bilinears, obtaining exact current dynamics beyond strict current-algebra closure. We also construct an exact Lindbladian in diagonal rational CFTs from Verlinde topological defect lines (TDLs), where modular data determine dephasing rates between distinguishable topological-charge sectors.

The remainder of this Letter proceeds as follows. After defining exact solvability via the closure of $\mathcal{L}^\dagger$ on an operator space, we show the exact solvability of Lindbladians in $N$ Majorana CFTs. Its solvability is then applied to $G_k$ WZW models via conformal Majorana embeddings. Finally, we construct the Verlinde-line Lindbladian, interpreting its exact dynamics as topological-charge dephasing, and conclude with a brief outlook.

\textit{Exact Solvability of the Lindbladian}.
In this letter, we formulate the concept of exact solvability within the Heisenberg picture. Under the action of a Lindblad superoperator $\cL$ (commonly referred to as the Lindbladian), the Heisenberg evolution of an operator $O$ is generated through
\begin{equation}
    \tr\left(O\mathcal{L}[\rho]\right)=\tr\left(\cL^\dagger[O]\rho\right)
\end{equation}
where $\rho$ is a normalized density matrix, so that an operator $O$ evolves as
\begin{equation}
\frac{\dd}{\dd t}O=\cL^\dagger[O]=\ii[H,O]
+\mathcal{D}^{\dagger}[O],
\label{eq:main-adjoint}
\end{equation}
where the dissipator is
\begin{equation}
    \mathcal{D}^{\dagger}[O]=\sum_i\left(K_i^\dagger O K_i
-\frac12\{K_i^\dagger K_i,O\}\right).
\end{equation}
Here, $H$ denotes the Hamiltonian and $K_i$ represent the jump operators. Following \cite{TangBaradWen2026}, we establish the notion of exact solvability. A Lindbladian is considered \textit{exactly solvable} within an operator space $\mathcal A$ if
\begin{equation}
\cL^\dagger[\mathcal A]\subseteq \mathcal A,
\label{eq:main-closure-principle}
\end{equation}
which is equivalent to the \textit{closure condition} of the dynamics \footnote{It is important to note that satisfying the closure condition does not constitute a full integrable Lindbladian. Consequently, it cannot determine the complete spectrum of $\cL$, nor all of its steady states or relaxation modes.}, expressed as
\begin{equation}
\cL^\dagger[O_\mu]=\sum_{\nu=1}^M M_{\mu\nu}O_\nu ,\quad \mathcal A=\operatorname{span}\{O_\mu\}_{\mu=1}^M
\label{eq:main-finite-matrix}.
\end{equation}

In what follows, we identify exactly solvable Lindbladians in $(1+1)$D CFTs with multiple Majorana fermions and generalize this approach to CFTs with affine currents using conformal embedding, for which a closure condition is secured. Furthermore, we construct an exact Lindbladian with Verlinde TDLs in $(1+1)$D CFTs corresponding to a dephasing process between topological charge sectors in $(2+1)$D topological phases.

\textit{Majorana CFTs}.
We first establish the basic solvable triangular hierarchy in an \(N\)-Majorana CFT. The Majorana modes obey the canonical anti-commutation relations (CARs)
\begin{equation}
\begin{gathered}
\{\psi_q^A,\psi_r^B\}
=
\delta^{AB}\delta_{q+r,0},
\qquad
(\psi_q^A)^\dagger=\psi_{-q}^A .
\end{gathered}
\label{eq:main-majorana-car}
\end{equation}
Here \(A=1,\ldots,N\) labels the Majorana species. For the chiral sector, the Hamiltonian is
\begin{equation}
H_\psi=
\frac{2\pi v}{L}
\left(L_0-\frac{c}{24}\right),
\qquad
c=\frac{N}{2},
\end{equation}
where
\begin{equation}
[L_0,\psi_q^A]=-q\psi_q^A .
\end{equation}
The anti-chiral sector is treated analogously. For a nonchiral theory one
would use
\[
H_{\psi,\mathrm{nonchiral}}
=
\frac{2\pi v}{L}
\left(L_0+\bar L_0-\frac{c}{12}\right).
\]
In what follows we work in the chiral sector unless stated otherwise. The mode index takes values \(q\in\mathbb Z+1/2\) in the Neveu-Schwarz sector and \(q\in\mathbb Z\) in the Ramond sector. Let \(\mathcal I^\ast=\mathcal I\setminus\{0\}\), where \(\mathcal I=\mathbb Z+1/2\) in the Neveu-Schwarz sector and \(\mathcal I=\mathbb Z\) in the Ramond sector. For every nonzero
Majorana mode \(q\in\mathcal I^\ast\), we introduce a Majorana-mode jump operator
\begin{equation}
K_q^A=\sqrt{\gamma(q,A)}\,\psi_q^A,
\qquad
\gamma(q,A)\ge0 .
\label{eq:main-majorana-jumps}
\end{equation}
For \(q>0\), \(K_q^A\) is a loss jump and \(K_{-q}^A\) is a gain jump.
For nonzero modes we define
\begin{equation}
\Gamma_{q,A}:=\gamma(q,A)+\gamma(-q,A),
\qquad
\Gamma_{q,A}=\Gamma_{-q,A}.
\end{equation}
In the Ramond sector, the zero modes satisfy
\begin{equation}
\psi_0^A=\psi_{-0}^A,
\qquad
(\psi_0^A)^2=\frac12,
\end{equation}
and should be treated as self-conjugate Clifford generators. All oscillator-pair formulas below are therefore stated for \(q>0\), while the zero-mode contribution is discussed separately at the end of this paragraph. This $N$ Majorana-mode Lindbladian is a CFT analog of quasi-free fermion Lindbladians \cite{Prosen2008NJP10,ProsenZunkovic2010NJP12,BarthelZhang2022JSTAT2022,Prosen2010JSTAT2010,ProsenSeligman2010Arxiv10072921}.

The adjoint Lindbladian acts as
\begin{equation}
\begin{split}
    \mathcal L_\psi^\dagger[O]
&=
i[H_\psi,O]
\\
&+
\sum_{A=1}^N\sum_{q\in\mathcal I^\ast}
\gamma(q,A)
\left(
\psi_{-q}^A O \psi_q^A
-\frac12\{\psi_{-q}^A\psi_q^A,O\}
\right).
\end{split}
\end{equation}
For nonzero modes \(p,r\in\mathcal I^\ast\), the CARs imply
\begin{equation}
\begin{aligned}
\frac{\dd}{\dd t}\left(\psi_p^A\psi_r^B\right)
&=
\left[
-i\frac{2\pi v}{L}(p+r)
-\frac{\Gamma_{p,A}}2
-\frac{\Gamma_{r,B}}2
\right]\psi_p^A\psi_r^B
\\
&\quad
+\delta^{AB}\delta_{p+r,0}\gamma(p,A)\,\mathbf{1} .
\end{aligned}
\label{eq:main-bilinear}
\end{equation}
Thus, when \(A\neq B\), or more generally when the source term vanishes, the solution is
\begin{equation}
\psi_p^A\psi_r^B(t)
=
e^{
\left[
-i\frac{2\pi v}{L}(p+r)
-\frac{\Gamma_{p,A}+\Gamma_{r,B}}2
\right]t}
\psi_p^A\psi_r^B(0).
\label{eq:majorana-bilinear-homogeneous-solution}
\end{equation}
For \(A=B\) and \(r=-p\), the equation contains an inhomogeneous source. For \(p\neq 0\),
\begin{equation}
\psi_{-p}^A\psi_p^A(t)
=
e^{-\Gamma_{p,A}t}\psi_{-p}^A\psi_p^A(0)
+
\frac{\gamma(-p,A)}{\Gamma_{p,A}}
\left(1-e^{-\Gamma_{p,A}t}\right)\mathbf{1} .
\label{eq:majorana-minus-p-p-solution}
\end{equation}

We then consider reduced ordered even monomials built from nonzero
Majorana modes,
\begin{equation}
O=
\psi_{p_1}^{A_1}
\psi_{p_2}^{A_2}
\cdots
\psi_{p_{2m}}^{A_{2m}},
\qquad
p_j\in\mathcal I^\ast,
\label{eq:Even-Maj-Monomial}
\end{equation}
where repeated nonzero modes have been simplified using the CARs.
Ramond zero modes are treated separately at the end of this paragraph. First suppose that \(O\) contains at most one member of every nonzero oscillator pair \((\psi_{-q}^A,\psi_q^A)\). Then each mode appearing in \(O\) only contributes its damping factor, the solution is given by
\begin{equation}
O(t)
=
e^{
\left[
-i\frac{2\pi v}{L}
\left(\sum_{j=1}^{2m}p_j\right)
-\frac12
\sum_{j=1}^{2m}
\Gamma_{p_j,A_j}
\right]t}
O(0).
\end{equation}

Now suppose that \(O\) contains both members of at least one nonzero oscillator pair. After anticommuting one such pair into adjacent positions, the monomial can be written as
\begin{equation}
O=\psi_{-q}^A\psi_q^A P,
\end{equation}
where \(P\) is an even product of Majorana modes from other oscillator pairs. Since \(\psi_{-q}^A\psi_q^A\) is even, it commutes with \(P\). The corresponding local dissipator is
\begin{equation}
\mathcal D_{qA}^{\dagger}
[\psi_{-q}^A\psi_q^A P]
=
-\Gamma_{q,A}\psi_{-q}^A\psi_q^A P
+
\gamma(-q,A)P .
\label{eq:Local-Maj-Dissipator-qA}
\end{equation}
For the opposite ordering one obtains
\begin{equation}
\mathcal D_{qA}^{\dagger}
[\psi_q^A\psi_{-q}^A P]
=
-\Gamma_{q,A}\psi_q^A\psi_{-q}^A P
+
\gamma(q,A)P .
\label{eq:Local-Maj-Dissipator-opposite-order}
\end{equation}
All other oscillator-pair dissipators act on \(P\) and then multiply the result by the displayed paired bilinear.

Equations~\eqref{eq:Local-Maj-Dissipator-qA} and \eqref{eq:Local-Maj-Dissipator-opposite-order} exhibit the recursive structure. The first term on the right hand side has the same Majorana degree as the original monomial, and the second term has degree lower by two. If the two paired modes are separated in the original ordering, the lower-degree term carries the fermionic sign generated by moving the two modes next to each other before contraction. More explicitly, if the original monomial contains
\[
\psi_{p_i}^{A}\cdots \psi_{p_j}^{A},
\qquad
p_j=-p_i,
\qquad
i<j,
\]
then deleting this pair produces the ordered monomial with the \(i\)-th and \(j\)-th entries removed, multiplied by the sign \((-1)^{j-i-1}\). The source rate is the rate associated with the left member of the contracted pair, namely \(\gamma(p_i,A)\).

Therefore \(\mathcal L_\psi^\dagger\) maps a degree-\(2m\) Majorana monomial to a linear combination of operators of degrees \(2m,2m-2,\ldots,2,0\). No higher-degree operators are generated. Hence the adjoint Lindbladian is triangular with respect to Majorana degree. Since the identity and bilinear sectors are solved explicitly, the four-Majorana sector is driven only by already solved bilinear and identity terms. Iterating this argument gives the exact Heisenberg dynamics of every finite reduced ordered even Majorana monomial.

Finally, in the Ramond sector the zero-mode dissipator is
\begin{equation}
\mathcal D_{0A}^\dagger[O]
=
\gamma(0,A)
\left(
\psi_0^A O\psi_0^A
-\frac12 O
\right),
\end{equation}
where we used \((\psi_0^A)^2=1/2\). For an ordered even monomial \(O\), one has \(\psi_0^A O\psi_0^A=\frac12 O\) if \(O\) does not contain \(\psi_0^A\), and \(\psi_0^A O\psi_0^A=-\frac12 O\) if \(O\) contains \(\psi_0^A\). Hence this dissipator either vanishes or gives \(-\gamma(0,A)O\). Thus the zero-mode dissipator is diagonal in the reduced monomial basis and leaves the triangular solvability unchanged. By contrast, odd monomials do not form a triangular closed sector under the full Majorana-mode Lindbladian. In what follows, we restrict our discussion to the Neveu-Schwarz sector; see the Supplemental Material (SM)~\cite{SM1} for a detailed discussion of the Ramond sector.

\paragraph{Exact current dynamics from conformal Majorana embeddings.}
WZW models provide canonical
realizations of affine Kac--Moody symmetries and play a central role in describing $(1+1)$D critical systems with continuous internal
symmetries. They form a natural setting for studying exactly solvable Lindbladian dynamics constrained by current algebra. However, as shown in \cite{TangBaradWen2026}, exact operator closure in current-mode Lindbladians is strongly restricted for non-Abelian \(G_k\) WZW models: generic products of affine currents do not close under the adjoint Lindbladian.

We overcome this restriction in conformal Majorana embeddings, where the \(G_k\) Sugawara stress tensor coincides with the stress tensor of an \(N\)-Majorana CFT \cite{10.1093/imrn/rnu092}. In such
embeddings, affine currents are realized as Majorana bilinears. Since linear Majorana-mode jump operators [Eq.~\eqref{eq:main-majorana-jumps}]
generate a triangular hierarchy on even Majorana monomials, products of currents inherit exact Heisenberg dynamics in the enlarged Majorana operator space, even when strict current-algebra closure fails.

The \(N\) Majorana fermions realize an ambient \(\mathrm{SO}(N)_1\) current algebra. When they transform in a real orthogonal representation of \(\mathfrak g=\mathrm{Lie}(G)\), their bilinears generate an affine \(\hat{\mathfrak g}_k\) subalgebra, which may be conformally embedded. For instance, \(\mathrm{SU}(2)_{10}\subset \mathrm{SO}(5)_1\), corresponding to five Majorana fermions \cite{PhysRevB.79.045316}. We choose \(t^a\) to be Hermitian, purely imaginary, and antisymmetric matrices
\begin{equation}
(t^a)^\dagger=t^a,\qquad (t^a)^T=-t^a,
\end{equation}
and they obey
\begin{equation}
[t^a,t^b]= i f^{ab}{}_{c}\,t^c ,
\end{equation}
where \(f^{ab}{}_{c}\) are the structure constants of \(\mathfrak g\).
The embedded affine currents are realized as normal-ordered Majorana
bilinears,
\begin{equation}
J_n^a=
\frac12
\sum_{A,B=1}^N\sum_{q\in\mathbb{Z}+\frac{1}{2}}
t^a_{AB}\normal{\psi_{n-q}^A\psi_q^B},
\label{eq:main-current-definition}
\end{equation}
which form the affine Kac--Moody
algebra \cite{DiFrancesco2012CFT}
\begin{equation}
[J_m^a,J_n^b]
=
\ii f^{ab}{}_{c}J_{m+n}^c
+
m\,\mathcal K^{ab}\delta_{m+n,0},
\label{eq:main-current-algebra}
\end{equation}
and the current action on Majorana modes is
\begin{equation}
[J_n^a,\psi_q^A]
=
-\sum_{B=1}^N t^a_{AB}\,\psi_{n+q}^B.
\label{eq:current-majorana-commutator}
\end{equation}
Here, the affine central matrix is fixed by the Majorana representation,
\begin{equation}
\mathcal K^{ab}
=
\frac12\Tr(t^a t^b).
\end{equation}
For a simple algebra, \(\mathcal K^{ab}=k\,\eta^{ab}\), where \(\eta^{ab}\)
is the basic invariant metric on \(\mathfrak g\).
In an
\(\eta\)-orthonormal basis, Eq.~\eqref{eq:main-current-algebra} reduces to
\begin{equation}
[J_m^a,J_n^b]
=
\ii f^{ab}{}_{c}J_{m+n}^c
+
k\,m\,\delta^{ab}\delta_{m+n,0}.
\end{equation}

We now restrict to conformal Majorana embeddings,
\begin{equation}
\widehat{\mathfrak g}_{k}\subset \widehat{\mathfrak{so}}(N)_{1},
\end{equation}
for which the Sugawara stress tensor of the embedded
\(\widehat{\mathfrak g}_{k}\) algebra coincides with the stress tensor of
the ambient Majorana theory. Equivalently, their chiral Hamiltonians match
\begin{equation}
H_\psi=H^{G_k}_{\rm WZW},
\end{equation}
and the central charges agree as well,
\begin{equation}
c_{\hat{\mathfrak g}_k}
=
\frac{k\,\dim\mathfrak g}{k+h^\vee}
=
\frac{N}{2},
\label{eq:conformal-embedding-central-charge}
\end{equation}
where \(h^\vee\) is the dual Coxeter number of \(\mathfrak g\). For a non-chiral theory with identical left- and right-moving embeddings, one has \(c=\bar{c}=N/2,~H_{\psi,\text{nonchiral}}=H^{G_k}_{\rm WZW,\text{nonchiral}}\).

In this conformal-embedding case, the Lindbladian constructed from linear
Majorana-mode jumps gives exact Heisenberg dynamics for the embedded
\(\hat{\mathfrak g}_k\) current algebra inside the ambient Majorana CFT. Thus the result
applies to WZW theories, or conformal-extension sectors, equipped with
this chosen Majorana realization. For generic mode- and
species-dependent jump rates, the evolution need not close within the
affine-current algebra itself. Nevertheless, because each current is a Majorana bilinear, the
dynamics remains exactly solvable in the closed triangular hierarchy of
Majorana monomials.\footnote{Not every WZW model admits a Majorana
realization. Moreover, if the embedding is not conformal, the Majorana
stress tensor decomposes as
\begin{equation}
T_\psi=T_{\hat{\mathfrak g}_k}^{\rm Sug}+T_{\rm coset}.
\end{equation}
The Majorana calculation then remains exact for the full ambient
Majorana theory governed by \(H_\psi\), but it cannot be identified with
pure \(G_k\) WZW dynamics unless the coset stress tensor vanishes,
or equivalently unless the central charges match.}

In the Neveu-Schwarz sector, the adjoint Lindbladian acts on a single affine current as
\begin{equation}
\begin{aligned}
\cL_\psi^\dagger[J_n^a]
&=-\ii\frac{2\pi v}{L}nJ_n^a
\\
&-\frac14\sum_{A,B=1}^N\sum_{q\in\mathbb{Z}+\frac{1}{2}}
t^a_{AB}\left(\Gamma_{n-q,A}+\Gamma_{q,B}\right)
\normal{\psi_{n-q}^A\psi_q^B}.
\end{aligned}
\label{eq:main-current-general}
\end{equation}
For generic rates, this equation is not closed in the strict current-algebra sense, since the second line cannot in general be written as a finite linear combination of affine currents. Nevertheless, \(J_n^a\) is mapped into the closed Majorana bilinear sector, and its dynamics is therefore exactly solvable through the triangular Majorana-mode hierarchy. A related exact single-current closure was obtained in \cite{TangBaradWen2026} under symmetric-rate conditions. In the present Majorana realization, imposing the uniform pair-rate condition
\begin{equation}
    \Gamma_{q,A}=\Gamma,~\forall q\in\mathcal{I}^*,~A\in\{1,2,\cdots N\},
\end{equation}
reduces Eq.~\eqref{eq:main-current-general} to
\begin{equation}
\cL_{\psi}^{\dagger}[J_n^a]=\left(-i\frac{2\pi v}{L}n-\Gamma\right)J_n^a,
\end{equation}
which closes even within the affine-current sector. The genuine challenge lies in multi-current dynamics, where current-algebra closure fails and the Majorana-bilinear embedding becomes essential for obtaining a closed triangular hierarchy.

Let us consider the Lindbladian dynamics of an ordered product of $m$ affine current modes
\begin{equation}
    O_m=J_{n_1}^{a_1}J_{n_2}^{a_2}\cdots J_{n_m}^{a_m}.
\end{equation}
Using Eqs.~\eqref{eq:main-majorana-car}, \eqref{eq:main-current-definition}, and \eqref{eq:main-current-algebra}, the adjoint Lindbladian admits the recursive form
\begin{equation}\label{eq:Affine-Maj-Lindbladian}
\begin{split}
    &\mathcal{L}_{\psi}^{\dagger}[O_m]=\mathcal{L}_{\psi}^{\dagger}[O_{m-1}]J_{n_m}^{a_m}-i\frac{2\pi v}{L}n_mO_m\\
    &-\frac{1}{4}O_{m-1}\sum_{A,B=1}^N\sum_{q\in\mathbb{Z}+\frac{1}{2}}t_{AB}^{a_m}\left[\Gamma_{(n_m-q),A}+\Gamma_{q,B}\right]:\psi_{n_m-q}^{A}\psi_{q}^{B}:\\
    &-\sum_{l=1}^{m-1}\sum_{C,D,E=1}^{N}\sum_{q\in\mathbb{Z}+\frac{1}{2}}\gamma(q,C)t^{a_l}_{CD}t^{a_m}_{CE}\\
    &\times J_{n_1}^{a_1}\cdots J_{n_{l-1}}^{a_{l-1}} \psi_{n_l-q}^{D} J_{n_{l+1}}^{a_{l+1}}\cdots J_{n_{m-1}}^{a_{m-1}} \psi_{n_m+q}^{E}.
\end{split}
\end{equation}
The first line of Eq.~\eqref{eq:Affine-Maj-Lindbladian} has maximal Majorana degree \(2m\): by induction, \(\mathcal{L}_{\psi}^{\dagger}[O_{m-1}]\) contains only even Majorana degrees not exceeding \(2m-2\), and multiplication by one current adds at most two Majorana modes. The second line contains \(m-1\) affine currents and one explicit Majorana bilinear, and hence also has maximal Majorana degree \(2m\) after expanding the currents using Eq.~\eqref{eq:main-current-definition}. Normal-ordering contractions may additionally generate lower-degree terms. The final summation contains \(m-2\) affine currents and two explicit Majorana modes, so after expanding the remaining currents it has maximal Majorana degree \(2m-2\), again with possible lower-degree contraction terms. Therefore, \(\mathcal{L}_{\psi}^{\dagger}[O_m]\) lies in the finite even Majorana hierarchy of degrees $2m,2m-2,\cdots 0$ and is exactly solvable, independently of the particular choice of non-negative Majorana jump rates \(\{\gamma(q,A)\}\). 

As a concrete illustration, we consider the ordered two-current operator
\begin{equation}
O_2=J_{n_1}^{a_1}J_{n_2}^{a_2}.
\end{equation}
Solving Eq.~\eqref{eq:Affine-Maj-Lindbladian} recursively gives an exact
expression for
\(O_2(t)\). The full result contains
Majorana monomials of degrees four, two, and zero, and is therefore
lengthy. We give its derivation and explicit form in the SM~\cite{SM1}. Here we present the long-time limit. Assuming that all
relevant pair rates \(\Gamma_{r,A}\) are nonzero, the degree-four and
degree-two terms decay, leaving only the double-contraction contribution,
\begin{equation}
\begin{split}
&\lim_{t\to\infty}
O_2(t)
\\
&\quad =
\frac{\delta_{n_1+n_2,0}}{2}
\sum_{p\in\mathbb{Z}+\frac{1}{2}}
\sum_{A,B=1}^{N}
t_{AB}^{a_1}t_{BA}^{a_2}
\,
\bar{\alpha}_{A,n_1-p}
\bar{\alpha}_{B,p}
\,\mathbf{1},
\end{split}
\label{eq:two-current-long-time-generic}
\end{equation}
where
\begin{equation}
\bar{\alpha}_{A,r}
=
\lim_{t\to\infty}\alpha_{A,r}(t)
=
\frac{\gamma(r,A)}{\Gamma_{r,A}}.
\end{equation}
For pure loss, $\gamma(r,A)>0,\gamma(-r,A)=0,r>0$, one has \(\bar{\alpha}_{A,r}=\Theta(r)\), where \(\Theta(r)=1\) for \(r>0\) and \(\Theta(r)=0\) for \(r<0\). Hence the long-time limit reduces to
\begin{equation}
\lim_{t\to\infty}
O_2(t)
=
k\,n_1\,\Theta(n_1)\,
\delta_{n_1+n_2,0}\delta^{a_1a_2}\,\mathbf{1}.
\label{eq:two-current-long-time-loss}
\end{equation}
Equivalently, for \(n>0\),
\begin{equation}
\lim_{t\to\infty}
\left[J_{n}^{a}J_{-n}^{b}\right](t)
=
k n\,\delta^{ab}\mathbf{1},
\qquad
\lim_{t\to\infty}
\left[J_{-n}^{a}J_{n}^{b}\right](t)
=
0.
\end{equation}
This result has a simple physical interpretation. Pure loss cools the Majorana oscillator modes to the vacuum functional, so the Heisenberg evolution projects a current product onto its vacuum contraction.

\textit{Exact CFT Lindbladian for topological-charge dephasing}.
Consider a diagonal rational CFT (RCFT), whose finitely many primary fields span the finite-dimensional primary-sector Hilbert space
\begin{equation}
\mathcal{H}_{p}
\cong
\mathrm{span}\left\{
\ket{a},\,a\in\mathcal{P}
\,\big|\,
L_0\ket{a}=\bar{L}_0\ket{a}=h_a\ket{a}
\right\},
\end{equation}
where \(h_a\) is the conformal dimension of the primary state \(\ket{a}\). Each primary \(x\in \mathcal{P}\) uniquely labels a Verlinde topological defect line (TDL) \(W_x\), which acts diagonally on \(\mathcal{H}_p\) as
\begin{equation}
     W_x=\sum_{a\in\mathcal{P}}\lambda_x(a)P_a,
    \qquad
    \lambda_x(a)=\frac{S_{xa}}{S_{\mathbf{1}a}}.
\end{equation}
Here \(x\) labels the Verlinde line, \(a\) labels the primary sector,
\(P_a=\ket{a}\bra{a}\), and \(S_{ab}\) are modular \(S\)-matrix elements.
In particular, $\lambda_x(\mathbf 1)=d_x$ is the quantum dimension of $W_x$. The Verlinde TDLs obey the closed fusion algebra
\begin{equation}
    W_aW_b=\sum_{x\in\mathcal{P}}N^x_{ab}W_x,
\end{equation}
where \(N^x_{ab}\) are the fusion coefficients.

We define the CFT Hamiltonian as
\begin{equation}
    H=\frac{2\pi v}{L}\left(L_0-\frac{c}{24}\right),
\end{equation}
where $H\ket{a}=\epsilon_a\ket{a}$. We take the Verlinde TDLs as Lindblad jump operators,
\begin{equation}
    K_x=\sqrt{\eta_x}W_x,\quad \eta_x\geq 0,
\end{equation}
whose Hermitian conjugates are determined by charge conjugation,
\begin{equation}
    (K_x)^\dagger=\sqrt{\eta_x}W_{\bar{x}} .
\end{equation}
The corresponding Heisenberg-picture Lindbladian is
\begin{equation}
    \mathcal{L}^{\dagger}_{V}[O]
    =
    i[H,O]
    +
    \sum_{x\in\mathcal{P}}
    \left(
    K_x^{\dagger}OK_x
    -
    \frac{1}{2}\{K_x^{\dagger}K_x,O\}
    \right).
\end{equation}

For an operator \(O\) acting within \(\mathcal{H}_p\),
\begin{equation}
    O=\sum_{a,b\in\mathcal{P}}O_{ab}P_{ab},
    \quad
    P_{ab}=\ket{a}\bra{b},
\end{equation}
with complex coefficients \(O_{ab}\), the Lindbladian dynamics closes within the primary-sector operator space and is therefore exactly solvable. The solution is
\begin{equation}
\label{eq:Exact-Primary-Solution-Lindbladian}
\begin{aligned}
O(t)
=
\sum_{a,b\in\mathcal P}
O_{ab}
\exp\!\left[
(-\Gamma_{ab}+i\omega_{ab})t
\right]
P_{ab},
\end{aligned}
\end{equation}
where the decay and oscillating factors are given by $\Gamma_{ab}
=
\frac12
\sum_{x\in\mathcal P}
\eta_x
|\lambda_x(a)-\lambda_x(b)|^2$, and $\omega_{ab}
=
\epsilon_a-\epsilon_b
+
\sum_{x\in\mathcal P}
\eta_x\,
\operatorname{Im}
\!\left[
\lambda_x(a)^*\lambda_x(b)
\right],$ respectively.
When the active Verlinde lines distinguish all primary sectors, the factor \(e^{-\Gamma_{ab}t}\) decays exponentially. Hence the infinite-time limit retains only the diagonal part,
\begin{equation}
\label{eq:TDL-Dephase}
    \lim_{t\to\infty}O(t)=\sum_{a\in\mathcal{P}}O_{aa}P_{a}.
\end{equation}
Thus the Lindbladian dephases coherences between distinct primary sectors while preserving the diagonal sector weights.

It is natural to ask for the time scale on which Eq.~\eqref{eq:TDL-Dephase} is reached, which is the dephasing time. It measures how fast the off-diagonal primary-sector coherences \(P_{ab}\), with \(a\neq b\), decay. From Eq.~\eqref{eq:Exact-Primary-Solution-Lindbladian}, the real decay rate of \(P_{ab}\) is \(\Gamma_{ab}\), which makes the distinction
between invertible and noninvertible TDLs transparent. For an invertible line \(x\), the eigenvalues \(\lambda_x(a)\) are U\((1)\) phases, so its single-line contribution is bounded by \(2\eta_x\). For a noninvertible line, \(\lambda_x(\mathbf 1)=d_x>1\) and the $\lambda_x(a),\lambda_x(b)$ need not have unit modulus, allowing a larger and more sector-selective dephasing rate. If the active jumps distinguish all primary sectors, then \(\Gamma_{ab}>0\) for every \(a\neq b\). The slowest distinguishable pair controls the longest dephasing time,
\begin{equation}
    \Gamma_V
    =
    \min_{a\neq b}\Gamma_{ab}>0.
\end{equation}
After a time of order \(\Gamma_V^{-1}\), all off-diagonal coherences are strongly suppressed, and \(O(t)\) is well approximated by its diagonal projection \(\sum_a O_{aa}P_a\). More precisely, for trace-distance accuracy \(\delta\in(0,1)\), the dephasing time satisfies
\begin{equation}
    t_{\mathrm{deph}}(\delta)
    \le\max\left\{0,
    \frac{1}{\Gamma_V}
    \log\left(
    \frac{\sqrt{|\mathcal P|}}{2\delta}
    \right)\right\}.
\end{equation}
The derivation of this bound and the corresponding trace-distance estimate are given in the SM~\cite{SM1}.

This exact dephasing process has a direct \((2+1)\)D bulk interpretation. In a diagonal RCFT that appears as the edge theory
of a \((2+1)\)D topological phase, the primary label \(a\) identifies a bulk
topological charge \cite{MooreSeiberg1989,Nayak2008}. The projector \(P_a\)
selects the corresponding topological-charge sector, and the Verlinde line
\(W_x\) acts as a topological Wilson-loop operator with eigenvalue
\(S_{xa}/S_{\mathbf 1 a}\)
\cite{Verlinde1988,PetkovaZuber2001,FrohlichFuchsRunkelSchweigert2007}. Hence
the jump \(K_x=\sqrt{\eta_x}W_x\) describes a continuous weak measurement of
topological charge by an external bath or detector \cite{JacobsSteck2006}.
Because \(W_x\) is diagonal in the topological-charge basis, the Lindbladian leaves the sector weights invariant and damps only the intersector coherences \(P_{ab}\) with \(a\neq b\). It thus realizes topological-charge dephasing, closely analogous to anyonic-charge decoherence in interferometric measurements
\cite{BondersonShtengelSlingerland2007,BondersonShtengelSlingerland2008}, dephasing by random anyonic environments \cite{ZatloukalLehmanSinghPachosBrennen2012}, and related decoherence phenomena in fractional quantum Hall and topologically ordered systems
\cite{ParkGefenSim2015,BaoFanVishwanathAltman2026,Hwang2024}. When the active lines resolve all sectors, every off-diagonal coherence decays while every diagonal weight remains fixed, making the Lindbladian an exactly solvable model of topological-charge dephasing. Its decay rates are fixed by the modular data and by the measurement strengths \(\eta_x\). This gives a direct link between edge RCFT, bulk topological-charge measurement, and open-system dynamics: the same modular \(S\) matrix that encodes braiding data also controls the exact decay of quantum coherence between topological-charge sectors.

\textit{Discussion}. In this Letter, we identified exactly tractable Lindbladian dynamics in CFT from an operator-closure viewpoint. The \(N\)-Majorana construction is a CFT realization of the quasi-free fermionic paradigm: linear Majorana jumps make the adjoint Lindbladian triangular on reduced even Majorana monomials, giving explicit recursive Heisenberg solutions. This further provides a solvable hierarchy that can be used beyond the free-fermion sector. In any $G_k$ WZW model admitting conformal Majorana embedding, affine currents realized as Majorana bilinears inherit this hierarchy, allowing exact affine current and multi-current dynamics even when closure within the current algebra itself is absent. We also constructed a distinct Verlinde-line Lindbladian in diagonal rational CFTs. The dynamics is exactly diagonal on operators connecting conformal-family sectors and realizes topological-sector dephasing with rates fixed by modular data and measurement strengths. Furthermore, we identify a family of exactly solvable Lindbladians within the
\(\operatorname{span}\{\mathbf 1,L_0,L_q,L_{-q}\}\) Virasoro $\text{SL}(2,\mathbb{R})$ subalgebra. Due to space constraints, we defer this discussion to the SM \cite{SM1}, and a more detailed treatment will be presented in \cite{UWW}.

Several directions remain open. Guided by the CFT analysis developed here and in Ref.~\cite{TangBaradWen2026}, exact Lindbladian dynamics may provide useful insight into steady-state properties of critical lattice systems. It would also be interesting to study dissipative ground-state preparation directly in CFT, a problem that has attracted recent attention in lattice systems but remains largely unexplored in conformal settings~\cite{Ding2024PRResearch6,Zhan2026PRX16,Lin2025Arxiv250521308}. In this Letter, we have worked mainly in the Heisenberg picture and focused on operator dynamics. A natural next question is whether comparable analytic control can be established at the level of states. Such a formulation could make steady-state properties directly accessible and may provide a useful framework for studying strong-to-weak symmetry breaking once symmetries are incorporated~\cite{LessaMaZhangBiChengWang2025PRXQuantum,
SalaGopalakrishnanOshikawaYou2024PRB,
KunoOritoIchinose2024PRB,
GuWangWang2025PRB,
HuangQiZhangLucas2025PRB}. We have also analyzed topological-charge dephasing within the subspace spanned by primary states. A further extension is to the finite-dimensional space of conformal blocks with several primary-field insertions, where additional categorical data, such as \(F\)- and \(R\)-symbols, become accessible. It would be interesting to determine what universal information can be extracted from exactly solvable Lindbladians that close within this conformal-block space.

\textit{Acknowledgment}. This work was supported by research funding provided to Masahiro Nozaki at the University of Chinese Academy of Sciences. C.B. thanks Xueda Wen for inspiring discussions and the encouragement to summarize these notes into this letter, and is grateful to both Xueda Wen and Qicheng Tang for their helpful comments on the manuscript. Furthermore, C.B. acknowledges the Croucher Foundation for supporting the Summer Course on ``Emergent Symmetry,'' where the initial ideas for this work were developed. The author acknowledges the assistance of ChatGPT 5.5 in verifying calculations and polishing the main text and Supplemental Material.

\bibliography{references}

\clearpage

\appendix

\begin{widetext}

\begin{center}
\textbf{Supplemental Material for ``Exact Lindbladian Dynamics from Conformal Embeddings and Topological Defects in Conformal Field Theory''}
\end{center}

\setcounter{table}{0}
\renewcommand{\thetable}{S\arabic{table}}
\setcounter{figure}{0}
\renewcommand{\thefigure}{S\arabic{figure}}
\setcounter{equation}{0}
\renewcommand{\theequation}{S\arabic{equation}}

This Supplemental Material provides detailed technical derivations supporting the results presented in the main text. In Sec.~I, we formulate the Majorana-mode Lindbladian for $N$-Majorana CFTs. We then derive the exact dynamics of Majorana bilinears and reduced even Majorana monomials, demonstrating the triangular closure of the even Majorana hierarchy. Additionally, we discuss the non-closure of odd monomials, the role of Ramond zero modes, pure Majorana cooling, the associated mixing-time bound, and finite-temperature detailed balance. In Sec.~II, we apply even Majorana hierarchy to affine currents in CFTs that admit conformal Majorana embeddings. We detail the Majorana realization of affine currents and derive the single-current dynamics, product rules, and exact multi-current recursion relations. Furthermore, as a simple non-trivial example, we study the exact finite- and long-time behavior for two currents and analyze the current decomposition in the Ramond sector. In Sec.~III, we construct the Verlinde-line Lindbladian in diagonal RCFTs, yielding its exact primary-sector solution that corresponds to a topological-charge sector dephasing. We evaluate the topological-charge dephasing rates, establish a distinguishability criterion and a trace-distance dephasing bound, and illustrate our framework with an explicit example in the Ising CFT. Finally, in Sec.~IV, we analyze an exactly closed Virasoro $\mathrm{SL}_q(2,\mathbb{R})$ block generated by $L_0, L_q$, and $L_{-q}$. We derive the exact symmetric-rate dynamics for the deformed Virasoro Hamiltonian using jump operators linear in $L_q$ and $L_{-q}$.

\tableofcontents

\section{Majorana-mode Lindbladian in CFT}
Throughout this section, we focus on the chiral sector. The anti-chiral sector is obtained analogously via the substitution $\psi \rightarrow \bar{\psi}$. For an $N$-component Majorana conformal field theory (CFT) defined on a circle of circumference $L$, the Majorana modes satisfy the canonical anticommutation relations (CARs):
\begin{equation}
\{\psi_q^A,\psi_r^B\}=\delta^{AB}\delta_{q+r,0},\qquad(\psi_q^A)^\dagger=\psi_{-q}^A,\qquad A=1,\ldots,N .\label{eq:SM-CAR}\end{equation}
The allowed momenta take values in the index set
\begin{equation}\mathcal{I}=\begin{cases}\mathbb{Z}+\frac{1}{2},\quad&\text{Neveu-Schwarz sector},\\
\mathbb{Z},\quad&\text{Ramond sector}.\end{cases}\end{equation}
We define the non-zero mode set as
\begin{equation}
\mathcal{I}^*=\mathcal{I}\setminus{\{0\}}.
\end{equation}
Thus, $\mathcal{I}^*$ excludes the Ramond zero mode and strictly contains pairs of finite-frequency oscillators. The chiral Hamiltonian $H_\psi$ and the central charge $c$ are given by
\begin{equation}
H_\psi=\frac{2\pi v}{L}\left(L_0-\frac{c}{24}\right),
\qquad
c=\frac N2 .
\label{eq:SM-Hpsi}
\end{equation}
where $v$ is a model-dependent constant and the Virasoro zero mode $L_0$ is defined as
\begin{equation}
L_0=\sum_{A=1}^{N}
\sum_{\substack{q\in\mathcal{I}^*,\ q>0}}
q\psi_{-q}^A\psi_q^A
+\text{constant}.
\end{equation}
Consequently, the Hamiltonian is a Majorana bilinear that satisfies the commutation relations
\begin{equation}
[L_0,\psi_q^A]=-q\psi_q^A,
\qquad
i[H_\psi,\psi_q^A]=-i\frac{2\pi v}{L}q\psi_q^A,\label{eq:SM-ladder}\end{equation}indicating that the $q>0$ modes act as lowering operators for $L_0$, and the $q<0$ modes act as raising operators. For each non-zero Majorana mode, we introduce a jump operator
\begin{equation}
K_q^A=\sqrt{\gamma(q,A)}\psi_q^A,
\qquad
q\in\mathcal{I}^*,
\qquad
\gamma(q,A)\ge 0 .
\label{eq:SM-Majorana-jump}
\end{equation}
For $q>0$, $K_q^A$ describes a loss process (since $\psi_q^A$ annihilates a positive-energy mode), whereas $K_{-q}^A$ describes a gain process (since $\psi_{-q}^A$ creates the corresponding mode). The total damping rate for this oscillator pair is defined as
\begin{equation}
\Gamma_{q,A}:=\gamma(q,A)+\gamma(-q,A),
\qquad
\Gamma_{q,A}=\Gamma_{-q,A}.
\label{eq:SM-Gamma}
\end{equation}
The corresponding adjoint Majorana Lindbladian $\mathcal{L}_\psi^\dagger$ is
\begin{equation}
\mathcal L_\psi^\dagger[O]=i[H_\psi,O]+\sum_{A=1}^{N}\sum_{q\in\mathcal{I}^*}\gamma(q,A)\left(\psi_{-q}^A O\psi_q^A-\frac12\{\psi_{-q}^A\psi_q^A,O\}\right).\label{eq:SM-adjoint-L}\end{equation}The first term governs the coherent unitary evolution of the CFT, while the second term captures the dissipation induced by all non-zero Majorana jump processes. The Heisenberg dynamics is closed, and thus exactly solvable, on a given operator subspace $\mathcal{A}$ if\begin{equation}\mathcal L_\psi^\dagger[\mathcal A]\subseteq \mathcal A .\label{eq:SM-closure}\end{equation}When this condition holds, the Heisenberg equations of motion do not generate operators outside of $\mathcal{A}$, and the time evolution within $\mathcal{A}$ reduces to a closed system of linear differential equations. Because the Hamiltonian in Eq.~\eqref{eq:SM-adjoint-L} is a Majorana bilinear and the jump operators are linear in the Majorana modes, this model serves as an exactly solvable CFT analog of a quasi-free fermionic Lindbladian \cite{Prosen2008NJP10,ProsenZunkovic2010NJP12,BarthelZhang2022JSTAT2022,Prosen2010JSTAT2010,ProsenSeligman2010Arxiv10072921}.

\subsection{Majorana Bilinears and higher even Majorana monomials}
In this part, we prove that any even monomial of Majorana modes exhibits exactly solvable, closed Lindbladian dynamics. First, we consider the bilinear
\begin{equation}
\psi_p^A\psi_r^B,
\qquad
p,r\in \mathcal{I}^* .
\end{equation}
The Hamiltonian contribution follows from Eq.~\eqref{eq:SM-ladder}
\begin{equation}
i[H_\psi,\psi_p^A\psi_r^B]
=
-i\frac{2\pi v}{L}(p+r)\psi_p^A\psi_r^B .
\label{eq:SM-bilinear-H}
\end{equation}
For an oscillator pair that does not appear in the bilinear, the corresponding local dissipator vanishes. Specifically, for an unpaired external pair $(q,C)$, we have
\begin{equation}
\psi_{-q}^C(\psi_p^A\psi_r^B)\psi_q^C
=
(\psi_p^A\psi_r^B)\psi_{-q}^C\psi_q^C,
\end{equation}
and
\begin{equation}
\frac12
\{\psi_{-q}^C\psi_q^C,\psi_p^A\psi_r^B\}
=
(\psi_p^A\psi_r^B)\psi_{-q}^C\psi_q^C .
\end{equation}
These two terms perfectly cancel, and the exact same cancellation holds for the gain jump. Consequently, the exact equation of motion for the bilinear is given by
\begin{equation}
\label{eq:SM-bilinear-eq}
\begin{split}
\mathcal L_\psi^\dagger[\psi_p^A\psi_r^B]
=&
\left[
-i\frac{2\pi v}{L}(p+r)
-\frac12\Gamma_{p,A}
-\frac12\Gamma_{r,B}
\right]\psi_p^A\psi_r^B
\\
&+
\delta^{AB}\delta_{p+r,0}\gamma(p,A)\,\mathbf 1 .
\end{split}
\end{equation}
which clearly closes within the subspace of bilinears and the identity. When the source term vanishes, the solution is
\begin{equation}
\psi_p^A\psi_r^B(t)
=
\exp\left\{
\left[
-i\frac{2\pi v}{L}(p+r)
-\frac12\Gamma_{p,A}
-\frac12\Gamma_{r,B}
\right]t
\right\}
\psi_p^A\psi_r^B(0).
\label{eq:SM-bilinear-homogeneous-sol}
\end{equation}
For $A=B$ and $r=-p>0$, with $\Gamma_{p,A}>0$, this becomes
\begin{equation}
\psi_{-p}^A\psi_p^A(t)
=
e^{-\Gamma_{p,A}t}\psi_{-p}^A\psi_p^A(0)
+
\frac{\gamma(-p,A)}{\Gamma_{p,A}}
\left(1-e^{-\Gamma_{p,A}t}\right)1 .
\label{eq:SM-occupation-general-p}
\end{equation}
If $\Gamma_{p,A}=0$, both rates vanish and the paired bilinear is exactly conserved.

Next, we consider a reduced, ordered even monomial constructed from nonzero Majorana modes,
\begin{equation}
O=
\psi_{p_1}^{A_1}
\psi_{p_2}^{A_2}
\cdots
\psi_{p_{2m}}^{A_{2m}},
\qquad
p_j\in \mathcal{I}^* .
\label{eq:SM-even-monomial}
\end{equation}
Here, ``reduced'' simply means that any repeated identical nonzero Majorana modes have already been eliminated using the CARs. If the same nonzero mode appears twice, the product automatically vanishes because
\begin{equation}
(\psi_p^A)^2=0,
\qquad
p\neq 0 .
\end{equation}
First, suppose that $O$ contains at most one member of each oscillator pair. In this case, no source term can appear, and each factor simply contributes its respective Hamiltonian phase and damping factor. Hence, the time evolution is
\begin{equation}
O(t)
=
\exp\left\{
\left[
-i\frac{2\pi v}{L}\sum_{j=1}^{2m}p_j
-\frac12\sum_{j=1}^{2m}\Gamma_{p_j,A_j}
\right]t
\right\}O(0).
\label{eq:SM-unpaired-even-sol}
\end{equation}
Now, suppose instead that $O$ contains both members of an oscillator pair. If the paired modes appear in adjacent positions as
\begin{equation}
\psi_{-q}^A\psi_q^A P,
\qquad
q>0,
\end{equation}
where $P$ is an even product built from the remaining oscillator pairs, then the dissipator acts as
\begin{equation}
D_{qA}^\dagger[\psi_{-q}^A\psi_q^A P]
=
-\Gamma_{q,A}\psi_{-q}^A\psi_q^A P
+
\gamma(-q,A)P .
\label{eq:SM-local-pair-minus-plus}
\end{equation}
For the opposite adjacent order,
\begin{equation}
D_{qA}^\dagger[\psi_q^A\psi_{-q}^A P]
=
-\Gamma_{q,A}\psi_q^A\psi_{-q}^A P
+
\gamma(q,A)P .
\label{eq:SM-local-pair-plus-minus}
\end{equation}
In both cases, the first term retains the exact same Majorana degree as the original monomial, while the second term has a degree reduced by two.

For a general ordered monomial, suppose the paired modes appear at positions $i<j$:
\begin{equation}
A_i=A_j,
\qquad
p_i+p_j=0 .
\end{equation}
Moving the $j$-th mode adjacent to the $i$-th mode requires commuting it past $j-i-1$ Majorana factors, which yields a fermion parity sign of $(-1)^{j-i-1}$. Let $O_{\widehat i\widehat j}$ denote the reduced ordered monomial obtained by deleting the $i$-th and $j$-th factors from $O$ but preserving the relative order of all remaining factors. We then obtain
\begin{align}
\mathcal L_\psi^\dagger[O]
=&
\left[
-i\frac{2\pi v}{L}\sum_{\ell=1}^{2m}p_\ell
-\frac12\sum_{\ell=1}^{2m}\Gamma_{p_\ell,A_\ell}
\right]O +\sum_{\substack{1\le i<j\le 2m\\ A_i=A_j,\\p_i+p_j=0}}
(-1)^{j-i-1}
\gamma(p_i,A_i)
O_{\widehat i\widehat j}.
\label{eq:SM-even-general-rule}
\end{align}
The source rate corresponds to the rate associated with the left member of the contracted pair. Equation~\eqref{eq:SM-even-general-rule} demonstrates that the dynamics are strictly triangular with respect to the Majorana degree. Specifically, a monomial of degree $2m$ maps exclusively to monomials of degrees
\begin{equation}
2m,\;2m-2,\;2m-4,\ldots,0 .
\label{eq:SM-degree-chain}
\end{equation}
No higher-degree operators are generated. Therefore, the even Majorana algebra is exactly solvable order by order in the degree. One can systematically solve the hierarchy by first finding the identity and the bilinear sector. The four-Majorana sector is then driven purely by these bilinears and the identity. Continuing this recursive procedure yields the exact Heisenberg dynamics for every finite, reduced, ordered even Majorana monomial.

\subsection{Odd monomials and Ramond zero modes}
In contrast to the even Majorana monomials, odd Majorana monomials generally do not form closed Lindbladian dynamics. Here, by analyzing the dynamics from a single mode up to general odd monomials, we demonstrate this non-closure, with the exception of specific single-mode cases. We first consider a single Majorana mode, $\psi_p^A$, which is odd under fermion parity. The Hamiltonian evolution is closed:
\begin{equation}
i[H_\psi,\psi_p^A]
=
-i\frac{2\pi v}{L}p\,\psi_p^A,
\label{eq:SM-single-H}
\end{equation}
and the loss jump gives
\begin{equation}
\psi_{-q}^A\psi_q^A\psi_q^A=0,
\qquad
\frac12\{\psi_{-q}^A\psi_q^A,\psi_q^A\}
=
\frac12\psi_q^A .
\end{equation}
Similarly, the gain jump gives
\begin{equation}
\psi_q^A\psi_q^A\psi_{-q}^A=0,
\qquad
\frac12\{\psi_q^A\psi_{-q}^A,\psi_q^A\}
=
\frac12\psi_q^A,
\end{equation}
such that the dissipator for the oscillator pair containing $\psi_p^A$ evaluates to
\begin{equation}
D_{|p|A}^\dagger[\psi_p^A]
=
-\frac12\Gamma_{p,A}\psi_p^A .
\label{eq:SM-single-own-pair}
\end{equation}
Now consider an external oscillator pair $(q,C)\neq(|p|,A)$ with $q>0$. The mode $\psi_p^A$ anticommutes with both $\psi_q^C$ and $\psi_{-q}^C$, and therefore commutes with the even products
\begin{equation}
\psi_{-q}^C\psi_q^C,
\qquad
\psi_q^C\psi_{-q}^C .
\end{equation}
For the external loss jump, we have
\begin{equation}
\psi_{-q}^C\psi_p^A\psi_q^C
=
-\psi_{-q}^C\psi_q^C\psi_p^A,\quad
\frac12
\{\psi_{-q}^C\psi_q^C,\psi_p^A\}
=
\psi_{-q}^C\psi_q^C\psi_p^A .
\end{equation}
Therefore this jump contributes
\begin{equation}
-2\gamma(q,C)\psi_{-q}^C\psi_q^C\psi_p^A .
\end{equation}
Similarly, the external gain jump contributes
\begin{equation}
-2\gamma(-q,C)\psi_q^C\psi_{-q}^C\psi_p^A .
\end{equation}
Combining all terms yields the exact equation of motion
\begin{align}
\mathcal L_\psi^\dagger[\psi_p^A]
=&
\left(
-i\frac{2\pi v}{L}p
-\frac12\Gamma_{p,A}
\right)\psi_p^A-
2
\sum_{C=1}^{N}
\sum_{\substack{q\in\mathcal{I}^*\\ q>0\\ (q,C)\neq(|p|,A)}}
\left[
\gamma(q,C)\psi_{-q}^C\psi_q^C
+
\gamma(-q,C)\psi_q^C\psi_{-q}^C
\right]\psi_p^A .
\label{eq:SM-single-majorana}
\end{align}
This equation illustrates why the degree-one odd sector is not closed. A single Majorana mode is mapped not only to itself but also to cubic odd operators, thereby generating higher-degree Majorana terms. If only the oscillator pair $(|p|,A)$ is coupled to the bath, all external terms in Eq.~\eqref{eq:SM-single-majorana} vanish, and one obtains a special exact solution for a single pair
\begin{equation}
\psi_p^A(t)
=
\exp\left[
\left(
-i\frac{2\pi v}{L}p
-\frac12\Gamma_{p,A}
\right)t
\right]\psi_p^A(0).
\label{eq:SM-single-special-sol}
\end{equation}

Next, we show that general odd monomials do not form the same finite-degree triangular hierarchy as their even counterparts. Under the Majorana-mode jump operators, external oscillator pairs generate higher-degree odd operators, meaning that a fixed, finite-degree odd subspace does not close. Let $O_{\rm odd}$ be an odd monomial that does not contain the oscillator pair $(\psi_q^A,\psi_{-q}^A)$, with $q>0$. The operator $O_{\rm odd}$ anticommutes with $\psi_q^A$ and $\psi_{-q}^A$, and therefore commutes with the bilinears $\psi_{-q}^A\psi_q^A$ and $\psi_q^A\psi_{-q}^A$. The local dissipator, comprising both loss and gain terms, gives
\begin{equation}
D_{qA}^\dagger[O_{\rm odd}]
=
-2\gamma(q,A)\psi_{-q}^A\psi_q^A O_{\rm odd}
-2\gamma(-q,A)\psi_q^A\psi_{-q}^A O_{\rm odd}.
\label{eq:SM-odd-external}
\end{equation}
Thus, an external oscillator pair adds two Majorana factors to an odd monomial, leading to a proliferation of Majorana modes in the Lindbladian dynamics.

\subsection{Zero Mode in Ramond Sector}
The Ramond sector requires additional discussion due to the presence of zero modes. These Ramond zero modes satisfy
\begin{equation}
\{\psi_0^A,\psi_0^B\}=\delta^{AB},
\qquad
(\psi_0^A)^\dagger=\psi_0^A,
\qquad
(\psi_0^A)^2=\frac12,
\label{eq:SM-zero-CAR}
\end{equation}
and generate a finite Clifford algebra. Furthermore, they commute with the CFT Hamiltonian. If a zero-mode jump operator
\begin{equation}
K_0^A=\sqrt{\gamma(0,A)}\,\psi_0^A ,
\end{equation}
is included, its adjoint dissipator, following from Eq.~\eqref{eq:SM-zero-CAR}, is given by
\begin{equation}
D_{0A}^\dagger[O]
=
\gamma(0,A)
\left(
\psi_0^A O\psi_0^A-\frac12 O
\right).
\label{eq:SM-zero-D}
\end{equation}

We divide our analysis into two distinct cases based on the structure of a reduced even monomial $O$. First, if $O$ does not contain $\psi_0^A$, it must commute with $\psi_0^A$. Consequently,
\begin{equation}
\psi_0^A O\psi_0^A=\frac12 O,
\qquad
D_{0A}^\dagger[O]=0 .
\label{eq:SM-zero-even-no-contain}
\end{equation}
Second, if $O$ does contain $\psi_0^A$, the remaining factors form an odd product that anticommutes with $\psi_0^A$. Thus,
\begin{equation}
\psi_0^A O\psi_0^A=-\frac12 O,
\qquad
D_{0A}^\dagger[O]=-\gamma(0,A)O .
\label{eq:SM-zero-even-contain}
\end{equation}
Therefore, the Ramond zero-mode jumps act purely diagonally in the basis of reduced even monomials. This ensures that they do not spoil the Majorana triangular hierarchy or the exact solvability of the system. If zero-mode jumps are explicitly excluded, the nonzero-mode jumps exactly conserve the even zero-mode algebra, which leads directly to the pure cooling protocol introduced below.

\subsection{Pure Majorana cooling and Finite Temperature Detailed Balance}
Here, we first study the pure cooling process for the $N$-Majorana CFT Lindbladian, which shows how fast the environment removes positive-energy Majorana excitations. Then, we extend our analysis to finite-temperature scenarios and the detailed-balance condition. We keep only the positive-mode loss jumps and set all gain rates to zero:
\begin{equation}
K_q^A=\sqrt{\gamma(q,A)}\,\psi_q^A,
\quad
\forall q>0,
\quad
\gamma(q,A)>0,\quad \gamma(-q,A)=0,
\label{eq:SM-cooling-jump}
\end{equation}
Since $\psi_q^A$ annihilates a positive-energy oscillator, this Lindbladian describes an environment that absorbs Majorana excitations from the CFT but does not inject them back.

Under Eq.~\eqref{eq:SM-cooling-jump}, Eq.~\eqref{eq:SM-occupation-general-p} reduces to
\begin{equation}
\psi_{-q}^A\psi_q^A(t)
=
e^{-\gamma(q,A)t}\psi_{-q}^A\psi_q^A(0),
\label{eq:SM-cooling-occupation}
\end{equation}
for every $q>0$. Thus, the occupation of each positive-energy oscillator decays exponentially with its own loss rate $\gamma(q,A)$. In the long-time limit,
\begin{equation}
\lim_{t\to\infty}\psi_{-q}^A\psi_q^A(t)=0,
\qquad
\lim_{t\to\infty}\psi_q^A\psi_{-q}^A(t)=1 .
\label{eq:SM-cooling-two-point}
\end{equation}
These are precisely the Neveu-Schwarz Fock-vacuum two-point expectation values. In this sense, the pure-loss Lindbladian cools the nonzero Majorana oscillator modes to the Neveu-Schwarz vacuum.

Let $|G\rangle$ denote the Neveu-Schwarz vacuum,
\begin{equation}
\psi_q^A|G\rangle=0,
\qquad
q>0 .
\end{equation}
To quantify the approach to this state, we define the total positive-mode occupation number
\begin{equation}
N_\psi
=
\sum_{A=1}^{N}
\sum_{\substack{q\in\mathcal I\\ q>0}}
\psi_{-q}^A\psi_q^A,
\label{eq:SM-Npsi}
\end{equation}
which counts the total number of nonzero Majorana excitations above the Neveu-Schwarz vacuum. We assume that the initial normalized density matrix $\rho(0)$ has a finite expected occupation, $\operatorname{tr}\!\left(N_\psi\rho(0)\right)<\infty$. This assumption excludes initial states with an infinite expected number of excited Majorana modes. The slowest decay rate among all positive modes is
\begin{equation}
\Delta_\psi
=
\inf_{\substack{q\in\mathcal I,\;q>0\\ A=1,\ldots,N}}
\gamma(q,A).
\label{eq:SM-Delta-psi}
\end{equation}
If $\Delta_\psi>0$, then every positive-mode occupation decays at least as fast as $e^{-\Delta_\psi t}$. Therefore, Eq.~\eqref{eq:SM-cooling-occupation} gives
\begin{equation}
\operatorname{tr}\!\left(N_\psi\rho(t)\right)
=
\operatorname{tr}\!\left(N_\psi(t)\rho(0)\right)
\le
e^{-\Delta_\psi t}
\operatorname{tr}\!\left(N_\psi\rho(0)\right).
\label{eq:SM-N-decay}
\end{equation}
Thus, $\Delta_\psi^{-1}$ sets the longest cooling time scale for the oscillator occupations. We now convert this occupation decay into a trace-distance bound. Since $N_\psi$ has an eigenvalue of zero only on the vacuum and has an eigenvalue of at least one on its orthogonal complement, one has the operator inequality
\begin{equation}
1-|G\rangle\langle G|\le N_\psi .
\label{eq:SM-vacuum-projector-bound}
\end{equation}
Therefore
\begin{equation}
1-\langle G|\rho(t)|G\rangle
=
\operatorname{tr}\!\left[(1-|G\rangle\langle G|)\rho(t)\right]
\le
\operatorname{tr}\!\left(N_\psi\rho(t)\right).
\label{eq:SM-fidelity-bound}
\end{equation}
The left-hand side represents the probability weight outside the target vacuum. Hence, Eq.~\eqref{eq:SM-fidelity-bound} indicates that the remaining distance from the vacuum is bounded by the remaining excitation number. We use the trace distance
\begin{equation}
D(\rho,\sigma)=\frac12\|\rho-\sigma\|_1 .
\end{equation}
for two arbitrary density matrices $\rho$ and $\sigma$, with $D(\rho,\sigma)\in[0,1]$. It measures how well the two states can be distinguished by measurements. Then, the Fuchs-van de Graaf inequality gives, for the pure target state $|G\rangle\langle G|$,
\begin{equation}
D(\rho(t),|G\rangle\langle G|)
\le
\sqrt{1-\langle G|\rho(t)|G\rangle}.
\end{equation}
Combining this inequality with Eqs.~\eqref{eq:SM-N-decay} and \eqref{eq:SM-fidelity-bound}, we obtain
\begin{equation}
D(\rho(t),|G\rangle\langle G|)
\le
\sqrt{\operatorname{tr}\!\left(N_\psi\rho(0)\right)}
\,e^{-\Delta_\psi t/2}.
\label{eq:SM-NS-trace-bound}
\end{equation}
The exponent is $\Delta_\psi/2$, rather than $\Delta_\psi$, because the trace-distance bound involves the square root of the excitation probability.

Let $0<\eta<1$ be a prescribed accuracy tolerance in trace distance. We define the mixing time $\tau_{\rm mix}(\eta)$ as the time after which the evolved state stays within a trace distance $\eta$ of the Neveu-Schwarz vacuum:
\begin{equation}
D(\rho(t),|G\rangle\langle G|)\le \eta
\qquad
\text{for all }t\ge \tau_{\rm mix}(\eta).
\end{equation}
Using Eq.~\eqref{eq:SM-NS-trace-bound}, one obtains the mixing-time upper bound
\begin{equation}
\tau_{\rm mix}(\eta)
\le
\max\left\{
0,\,
\frac{2}{\Delta_\psi}
\log
\left[
\frac{\sqrt{\operatorname{tr}(N_\psi\rho(0))}}{\eta}
\right]
\right\}.
\label{eq:SM-NS-mixing}
\end{equation}
The parameter $\eta$ is not a physical rate but a chosen tolerance. A smaller $\eta$ means that we demand the final state to be closer to the vacuum. The bound shows that the cooling time grows logarithmically with the initial expected number of excitations and also logarithmically with $1/\eta$. The prefactor is fixed by the inverse of the slowest loss rate, $\Delta_\psi^{-1}$.

In the Ramond sector, the same cooling jumps remove only positive nonzero modes but act trivially on the zero modes. Let $P_R$ denote the projector onto the (generally degenerate) subspace
\begin{equation}
\mathcal{H}_{R}
=
\left\{
|\Phi\rangle:
\psi_q^A|\Phi\rangle=0
\text{ for all }q>0,\;A=1,\ldots,N
\right\}.
\label{eq:SM-PR}
\end{equation}
which is annihilated by all positive nonzero modes. The same number-operator argument used for the Neveu-Schwarz sector yields
\begin{equation}
1-P_R\le N_\psi .
\label{eq:SM-PR-bound}
\end{equation}
Hence,
\begin{equation}
1-\operatorname{tr}(P_R\rho(t))
\le
e^{-\Delta_\psi t}
\operatorname{tr}\!\left(N_\psi\rho(0)\right).
\label{eq:SM-Ramond-cooling-bound}
\end{equation}
This is a cooling-to-subspace bound. It demonstrates that the nonzero oscillator excitations are removed exponentially, at the same time, the zero-mode density matrix remains invariant. Every density matrix supported in the Ramond nonzero-mode vacuum subspace $\mathcal{H}_R$ is stationary under the pure cooling dissipator
\begin{equation}
\rho_0=P_R\rho_0P_R
\quad\Longrightarrow\quad
\mathcal L_{\psi,{\rm cool}}[\rho_0]=0 .
\label{eq:SM-Ramond-stationary}
\end{equation}
A unique Ramond ground state requires an additional choice of the zero-mode sector.

We now demonstrate that introducing both loss and gain jumps causes the Majorana oscillator to relax toward a finite-temperature Gibbs state. Fix $q>0$ and $A$, the positive oscillator has energy
\begin{equation}
E_q=\frac{2\pi v q}{L}.
\end{equation}
The two local states are the empty state and the occupied state. Their respective projectors are
\begin{equation}
\psi_q^A\psi_{-q}^A,
\qquad
\psi_{-q}^A\psi_q^A,
\end{equation}
respectively. At inverse temperature $\beta$, the local Gibbs density matrix of this oscillator pair is
\begin{equation}
\rho_{\beta,qA}
=
\frac{
\psi_q^A\psi_{-q}^A
+
e^{-\beta E_q}\psi_{-q}^A\psi_q^A
}{
1+e^{-\beta E_q}
}.
\label{eq:SM-local-Gibbs}
\end{equation}
The occupied state is suppressed by the Boltzmann factor $e^{-\beta E_q}$. Therefore, the thermal occupation is
\begin{equation}
\operatorname{tr}\!\left[
\psi_{-q}^A\psi_q^A\rho_{\beta,qA}
\right]
=
\frac{1}{e^{\beta E_q}+1},
\label{eq:SM-FD}
\end{equation}
which is exactly the Fermi-Dirac distribution for one fermionic oscillator.

The local Schr\"odinger-picture dissipator is
\begin{align}
D_{qA}[\rho]
=\gamma(q,A)
\left(
\psi_q^A\rho\psi_{-q}^A
-\frac12\{\psi_{-q}^A\psi_q^A,\rho\}
\right)+\gamma(-q,A)
\left(
\psi_{-q}^A\rho\psi_q^A
-\frac12\{\psi_q^A\psi_{-q}^A,\rho\}
\right).
\label{eq:SM-Sch-local-D}
\end{align}
The first term is the loss channel, which removes an occupied fermion and sends the occupied state to the empty state. The second term is the gain channel, which injects a fermion and sends the empty state to the occupied state. Using the CARs, the loss channel acts nontrivially only on the occupied projector $\psi_{-q}^A\psi_q^A
\rightarrow
\psi_q^A\psi_{-q}^A-\psi_{-q}^A\psi_q^A$. Similarly, the gain channel acts nontrivially only on the empty projector $\psi_q^A\psi_{-q}^A
\rightarrow
\psi_{-q}^A\psi_q^A-\psi_q^A\psi_{-q}^A$.
Applying these two actions to Eq.~\eqref{eq:SM-local-Gibbs}, one obtains
\begin{align}
D_{qA}[\rho_{\beta,qA}]
=
\frac{
\gamma(q,A)e^{-\beta E_q}-\gamma(-q,A)
}{
1+e^{-\beta E_q}
}
\left(
\psi_q^A\psi_{-q}^A-\psi_{-q}^A\psi_q^A
\right).
\label{eq:SM-thermal-local-action}
\end{align}
Hence, the local Gibbs density matrix is stationary if and only if the detailed-balance condition is satisfied
\begin{equation}
\frac{\gamma(-q,A)}{\gamma(q,A)}
=
e^{-\beta E_q}
=
\exp\left(-\beta\frac{2\pi v q}{L}\right),
\qquad
q>0 .
\label{eq:SM-detailed-balance}
\end{equation}
Physically, this means that the bath can both absorb and supply energy, but upward transitions are thermally suppressed. The same condition follows from the Heisenberg occupation equation. From Eq.~\eqref{eq:SM-occupation-general-p}, the stationary value of the occupation satisfies
\begin{equation}
\frac{\gamma(-q,A)}{\gamma(q,A)+\gamma(-q,A)}
=
\frac{1}{e^{\beta E_q}+1},
\label{eq:SM-stationary-FD}
\end{equation}
when Eq.~\eqref{eq:SM-detailed-balance} holds. Thus, the Lindbladian steady point precisely exhibits the Fermi-Dirac occupation of the Majorana oscillator. For a fixed spin sector, the full finite-temperature state is the product of the local factors $\rho_{\beta,qA}$ over all $q>0$ and all species $A$. Since each oscillator pair is stationary under its own dissipator, and since $H_\psi$ is diagonal in the same occupation basis, the full Gibbs state satisfies
\begin{equation}
\mathcal L_\psi[\rho_\beta]=0
\label{eq:SM-thermal-stationary}
\end{equation}
under the detailed-balance condition in Eq.~\eqref{eq:SM-detailed-balance}. Therefore, the same Majorana-mode Lindbladian can describe either zero-temperature cooling or finite-temperature relaxation, depending on the ratio between gain and loss rates. Pure cooling is recovered as the zero-temperature limit. Keeping the loss rate $\gamma(q,A)>0$ fixed, Eq.~\eqref{eq:SM-detailed-balance} gives
\begin{equation}
\lim_{\beta\to\infty}\gamma(-q,A)
=
\lim_{\beta\to\infty}
\gamma(q,A)e^{-\beta E_q}
=
0,
\qquad
q>0 .
\end{equation}
At the same time,
\begin{equation}
\lim_{\beta\to\infty}
\frac{1}{e^{\beta E_q}+1}
=
0 .
\end{equation}
Thus, the zero-temperature limit removes all gain processes and leaves only positive-mode loss jumps.

In the Ramond sector, the nonzero modes obey the same detailed-balance relation. The only new feature is that the Ramond zero modes have zero energy, so their detailed-balance factor is
\begin{equation}
e^{-\beta(2\pi v\cdot 0/L)}=1 .
\end{equation}
Thus, finite temperature alone does not uniquely select a zero-mode occupation based on energy. If no zero-mode jumps are included, any even zero-mode density matrix can be tensored with the nonzero-mode thermal product; conversely, if zero-mode jumps are included, the canonical zero-mode identity state becomes stationary. Therefore, detailed balance fixes the thermal occupations of all nonzero oscillator modes, while the Ramond zero-mode sector remains unchanged unless an additional zero-mode dissipator is specified.

\section{Exact current dynamics from conformal Majorana embeddings}
\label{sec:SM-current-embedding}
Having established that the Majorana-mode Lindbladian exhibits a triangular hierarchy leading to exact solvability, we now demonstrate that this hierarchy can be applied to affine Kac-Moody currents in interacting CFTs, provided the current algebra is realized by Majorana bilinears. In such theories, each affine current is an even Majorana bilinear, and any product of $m$ currents is an even Majorana monomial with a maximal Majorana degree of $2m$. Consequently, due to the triangular hierarchy, its Heisenberg evolution remains confined within the closed even Majorana algebra and is exactly solvable.

It is important to note that this does not imply the current sector consists solely of free-fermion observables. From a current-algebra perspective, these sectors can describe interacting Wess-Zumino-Witten (WZW) conformal fixed points—such as the $\text{SU}(2)_{10}$ model embedded into a CFT of five Majorana fermions \cite{PhysRevB.79.045316}. The conformal Majorana embedding thus provides an exact method for solving the current dynamics by enlarging the operator space to encompass the full even Majorana algebra.

\subsection{Conformal Embedding and Majorana Realization of Affine Currents}
\label{subsec:SM-current-realization}

Let $\hat{g}_k$ be the compact affine Lie algebra at level $k$ associated with the Lie group $G$, and let the Majorana fermions transform in a real orthogonal representation of the finite-dimensional Lie algebra $g = \mathrm{Lie}(G)$. We choose Hermitian matrices $t^a$, with $a=1,\ldots,\dim g$, satisfying
\begin{equation}
[t^a,t^b]= i f^{ab}{}_{c} t^c .
\end{equation}
Since the representation is real orthogonal, the matrices are antisymmetric in the Majorana species labels
\begin{equation}
t^a_{AB}=-t^a_{BA}.
\end{equation}
Then, the affine currents are realized as Majorana bilinears,
\begin{equation}
J_n^a
=
\frac12
\sum_{A,B=1}^N
\sum_{q\in\mathcal I}
t^a_{AB}
:\psi^A_{n-q}\psi^B_q: .
\label{eq:SM-current-Majorana-realization}
\end{equation}
Here, $n\in\mathbb{Z}$ and $q\in\mathcal{I}^*$ label the current and Majorana modes, respectively. These affine currents generate the Kac-Moody algebra
\begin{equation}
[J_m^a,J_n^b]
=
i f^{ab}{}_{c}J_{m+n}^c
+
k\,m\,\delta^{ab}\delta_{m+n,0},
\label{eq:SM-affine-algebra}
\end{equation}
and the action of the current on a Majorana mode is given by
\begin{equation}
[J_n^a,\psi_q^A]
=
-\sum_{B=1}^N t^a_{AB}\psi^B_{n+q}.
\label{eq:SM-current-action-majorana}
\end{equation}

This construction applies to affine current algebras $\hat{g}_k$ that admit such a Majorana-bilinear realization. To identify the Majorana Hamiltonian with the pure $G_k$ WZW Hamiltonian, we impose the conformal-embedding condition
\begin{equation}
H_\psi=H_{\rm WZW}^{G_k}.
\label{eq:SM-conformal-embedding-H}
\end{equation}
Equivalently, the Sugawara central charge of the embedded $\hat{g}_k$ current algebra must equal the Majorana central charge
\begin{equation}
c_{\hat{g}_k}
=
\frac{k\,\dim g}{k+h^\vee}
=
\frac{N}{2}.
\label{eq:SM-conformal-embedding-c}
\end{equation}
Here, $h^\vee$ is the dual Coxeter number of the finite Lie algebra $g$.

For a nonchiral theory, there is also an anti-chiral sector. We denote the anti-chiral Majorana modes by $\bar{\psi}_q^A$, and the anti-chiral currents by $\bar{J}_n^a$. They obey the same equations as the holomorphic sector, with the substitutions
\begin{equation}
\psi\rightarrow \bar\psi,
\qquad
J\rightarrow \bar J,
\qquad
\gamma\rightarrow \bar\gamma .
\end{equation}
If the left- and right-moving sectors both share the same conformal Majorana embedding, then
\begin{equation}
c_{\hat{g}_k}=\bar c_{\hat{g}_k}=\frac{N}{2},
\end{equation}
and the nonchiral Hamiltonian is
\begin{equation}
H_{\psi,\mathrm{nonchiral}}
=
H_\psi+\bar H_\psi
=
H_{\rm WZW}^{G_k}
+
\bar H_{\rm WZW}^{G_k}.
\end{equation}
The holomorphic and anti-holomorphic current algebras commute. Therefore, for a purely holomorphic operator, the anti-chiral Lindbladian does not contribute. For an operator containing both $J$ and $\bar{J}$, the full adjoint Lindbladian is the sum of the two sector contributions, and the same Majorana hierarchy applies separately within each sector.

If the embedding is not conformal, the Majorana hierarchy remains exact within the ambient Majorana theory. However, the stress tensor then decomposes as
\begin{equation}
T_\psi=T^{\rm Sug}_{\hat{g}_k}+T_{\rm coset}.
\end{equation}
In that case, the dynamics generated by $H_\psi$ do not correspond to pure $G_k$ WZW dynamics unless the coset stress tensor vanishes. For the purposes of this work, we restrict our discussion entirely to conformal embeddings.

\subsection{Single-current dynamics}
\label{subsec:SM-single-current}

We now compute the action of $\mathcal L_\psi^\dagger$ on a single current $J_n^a$. Throughout this section, we work in the Neveu-Schwarz sector unless otherwise specified; the Ramond sector is discussed independently later. By inserting Eq.~\eqref{eq:SM-bilinear-eq} into the bilinear definition of $J_n^a$, and noting that every bilinear in $J_n^a$ has a total mode number of $(n-q)+q=n$, the Hamiltonian contribution evaluates to $-i\frac{2\pi v}{L}nJ_n^a$. The source term in Eq.~\eqref{eq:SM-bilinear-eq} arises only when the two Majorana modes are conjugate and share the same species label. In the current sum, this source term is multiplied by $t^a_{AA}=0$ due to the antisymmetry of $t^a$, yielding
\begin{equation}
\begin{aligned}
\mathcal L_\psi^\dagger[J_n^a]
=&
-i\frac{2\pi v}{L}nJ_n^a-
\frac14
\sum_{A,B=1}^N
\sum_{q\in\mathcal I}
t^a_{AB}
\left(
\Gamma_{n-q,A}
+
\Gamma_{q,B}
\right)
:\psi^A_{n-q}\psi^B_q:.
\end{aligned}
\label{eq:SM-single-current-general}
\end{equation}
For general mode- and species-dependent decay rates, the right-hand side is a weighted sum of Majorana bilinears that does not generally reduce to a finite linear combination of currents. Thus, the current subspace does not close under arbitrary rates. Nevertheless, the result remains entirely within the closed Majorana bilinear sector. Therefore, the time evolution of a single current remains exactly solvable by virtue of the conformal Majorana embedding.

Strict closure of the current sector occurs only when the total decay rate is uniform, meaning
\begin{equation}
\Gamma_{q,A}=\Gamma
\qquad
\text{for all }q\in\mathcal I,\; A=1,\ldots,N .
\label{eq:SM-uniform-total-rate}
\end{equation}
In this uniform limit, the dissipative term simplifies to $-\Gamma J_n^a$, leading to
\begin{equation}
\mathcal L_\psi^\dagger[J_n^a]
=
\left(
-i\frac{2\pi v}{L}n-\Gamma
\right)J_n^a,
\label{eq:SM-single-current-closed}
\end{equation}
which admits the exact solution
\begin{equation}
J_n^a(t)
=
\exp\left[
\left(
-i\frac{2\pi v}{L}n-\Gamma
\right)t
\right]J_n^a(0).
\label{eq:SM-single-current-solution}
\end{equation}

\subsection{Product rule and multi-current recursion}
\label{subsec:SM-current-product-rule}
For two operators $X$ and $Y$, the adjoint Lindbladian obeys the product rule
\begin{equation}
\begin{aligned}
\mathcal L_\psi^\dagger[XY]
=&
\mathcal L_\psi^\dagger[X]Y
+
X\mathcal L_\psi^\dagger[Y]+
\sum_{C=1}^N
\sum_{q\in\mathcal I}
\gamma(q,C)
[\psi^C_{-q},X][Y,\psi^C_q].
\end{aligned}
\label{eq:SM-product-rule}
\end{equation}
To prove this, let $\left. \cdots \right|_{K_i}$ denote the contribution originating from a single jump operator $K_i$. Evaluating the left-hand side yields
\begin{equation}
\begin{aligned}
\left.\mathcal L_\psi^\dagger[XY]\right|_{K_i}
&=
K_i^\dagger XYK_i-\frac12\{K_i^\dagger K_i,XY\}=
K_i^\dagger XYK_i
-\frac12K_i^\dagger K_iXY
-\frac12XYK_i^\dagger K_i .
\end{aligned}
\label{eq:SM-product-rule-LHS-one-jump}
\end{equation}
The first term on the right-hand side of Eq.~\eqref{eq:SM-product-rule} evaluates to
\begin{equation}
\begin{aligned}
\left.\mathcal L_\psi^\dagger[X]Y\right|_{K_i}
&=
\left(
K_i^\dagger XK_i-\frac12\{K_i^\dagger K_i,X\}
\right)Y=
K_i^\dagger XK_iY
-\frac12K_i^\dagger K_iXY
-\frac12XK_i^\dagger K_iY .
\end{aligned}
\end{equation}
The second term gives
\begin{equation}
\begin{aligned}
\left.X\mathcal L_\psi^\dagger[Y]\right|_{K_i}
&=
X\left(
K_i^\dagger YK_i-\frac12\{K_i^\dagger K_i,Y\}
\right)=
XK_i^\dagger YK_i
-\frac12XK_i^\dagger K_iY
-\frac12XYK_i^\dagger K_i .
\end{aligned}
\end{equation}
The correction commutator term yields
\begin{equation}
\begin{aligned}
[K_i^\dagger,X][Y,K_i]
&=
(K_i^\dagger X-XK_i^\dagger)(YK_i-K_iY)=
K_i^\dagger XYK_i
-K_i^\dagger XK_iY
-XK_i^\dagger YK_i
+XK_i^\dagger K_iY .
\end{aligned}
\end{equation}
Summing the last three equations gives
\begin{equation}
\begin{aligned}
&\left.\mathcal L_\psi^\dagger[X]Y\right|_{K_i}
+
\left.X\mathcal L_\psi^\dagger[Y]\right|_{K_i}
+
[K_i^\dagger,X][Y,K_i]
\\
&=
K_i^\dagger XK_iY
-\frac12K_i^\dagger K_iXY
-\frac12XK_i^\dagger K_iY+
XK_i^\dagger YK_i
-\frac12XK_i^\dagger K_iY
-\frac12XYK_i^\dagger K_i
\\
&\quad+
K_i^\dagger XYK_i
-K_i^\dagger XK_iY
-XK_i^\dagger YK_i
+XK_i^\dagger K_iY
\\
&=
K_i^\dagger XYK_i
-\frac12K_i^\dagger K_iXY
-\frac12XYK_i^\dagger K_i=\left.\mathcal L_\psi^\dagger[XY]\right|_{K_i}.
\end{aligned}
\end{equation}
Thus, the additional commutator term exactly accounts for the correction to the naive product rule for a single jump. For the Majorana jumps
\begin{equation}
K_q^C=\sqrt{\gamma(q,C)}\psi_q^C,
\qquad
(K_q^C)^\dagger=\sqrt{\gamma(q,C)}\psi_{-q}^C,
\end{equation}
this correction becomes
\begin{equation}
[(K_q^C)^\dagger,X][Y,K_q^C]
=
\gamma(q,C)[\psi_{-q}^C,X][Y,\psi_q^C].
\end{equation}
After summing over $C$ and $q$ and incorporating the Hamiltonian identity
\begin{equation}
i[H_\psi,XY]
=
i[H_\psi,X]Y+iX[H_\psi,Y],
\end{equation}
we successfully obtain Eq.~\eqref{eq:SM-product-rule}.

Next, we consider products of affine currents
\begin{equation}
O_m
=
J_{n_1}^{a_1}
J_{n_2}^{a_2}
\cdots
J_{n_m}^{a_m},\quad O_{m-1}
=
J_{n_1}^{a_1}\cdots J_{n_{m-1}}^{a_{m-1}},
\end{equation}
and apply Eq.~\eqref{eq:SM-product-rule} using
\begin{equation}
X=J_{n_1}^{a_1}\cdots J_{n_{m-1}}^{a_{m-1}},
\qquad
Y=J_{n_m}^{a_m}.
\end{equation}
The cross-commutator term is evaluated using
\begin{equation}
[\psi^C_{-q},J_{n_\ell}^{a_\ell}]
=
\sum_D t^{a_\ell}_{CD}\psi^D_{n_\ell-q},
\qquad
[J_{n_m}^{a_m},\psi_q^C]
=
-\sum_E t^{a_m}_{CE}\psi^E_{n_m+q}.
\end{equation}
Substituting the arbitrary-rate single-current result from Eq.~\eqref{eq:SM-single-current-general}, we obtain the exact recursion relation
\begin{equation}
\begin{aligned}
\mathcal L_\psi^\dagger[O_m]
=&\;
\mathcal L_\psi^\dagger[O_{m-1}]J_{n_m}^{a_m}
-i\frac{2\pi v}{L}n_m O_m-
\frac14
O_{m-1}
\sum_{A,B=1}^N
\sum_{q\in\mathcal I}
t^{a_m}_{AB}
\left(
\Gamma_{n_m-q,A}
+
\Gamma_{q,B}
\right)
:\psi^A_{n_m-q}\psi^B_q:
\\
&-
\sum_{\ell=1}^{m-1}
\sum_{C,D,E=1}^N
\sum_{q\in\mathcal I}
\gamma(q,C)
t^{a_\ell}_{CD}t^{a_m}_{CE}J_{n_1}^{a_1}\cdots
J_{n_{\ell-1}}^{a_{\ell-1}}
\psi^D_{n_\ell-q}
J_{n_{\ell+1}}^{a_{\ell+1}}
\cdots
J_{n_{m-1}}^{a_{m-1}}
\psi^E_{n_m+q},
\end{aligned}
\label{eq:SM-multicurrent-recursion}
\end{equation}
which is exactly solvable for arbitrary nonnegative decay rates $\gamma(q,A)$. The closure of this equation can be verified straightforwardly by counting the Majorana degree. Because each current is a Majorana bilinear, $O_m$ has a maximal Majorana degree of $2m$. The first term in Eq.~\eqref{eq:SM-multicurrent-recursion} contains $\mathcal L_\psi^\dagger[O_{m-1}]$ multiplied by one additional current. By induction, $\mathcal L_\psi^\dagger[O_{m-1}]$ contains only even Majorana degrees not exceeding $2m-2$, and multiplying by $J_{n_m}^{a_m}$ raises the maximal degree by at most two. Hence, this term has a degree no larger than $2m$. The Hamiltonian term proportional to $O_m$ also has a maximal degree of $2m$. The arbitrary-rate single-current dissipative term is $O_{m-1}$ multiplied by a weighted Majorana bilinear, thus also achieving a maximal degree of $2(m-1)+2=2m$. The final term contains $m-2$ currents and two explicit Majorana modes. After expanding the remaining currents, its maximal degree is $2(m-2)+2=2m-2$. Note that normal-ordering contractions can only lower the Majorana degree by even integers. Therefore, $\mathcal L_\psi^\dagger[O_m]$ is strictly confined to the finite even Majorana triangular hierarchy of degrees
\begin{equation}
2m,\;2m-2,\;\ldots,\;0 .
\end{equation}

In a non-Abelian affine current algebra, products of currents typically generate additional current structures and shifted modes, which can cause the current algebra alone to fail to close under dissipation \cite{TangBaradWen2026}. Conformal embedding provides a larger closed operator space. Within this extended space, the dissipator either preserves the even Majorana degree or contracts a pair to lower the degree by two, but it never increases the degree. This mechanism guarantees the exact solvability of the multi-current dynamics.
We now illustrate this with the simplest nontrivial example: a product of two affine currents,
\begin{equation}
O_2=J_{n_1}^{a_1}J_{n_2}^{a_2}.
\end{equation}
Using Eq.~\eqref{eq:SM-product-rule}, we obtain
\begin{equation}
\begin{aligned}
\mathcal L_\psi^\dagger[
J_{n_1}^{a_1}J_{n_2}^{a_2}
]
=&\;
\mathcal L_\psi^\dagger[J_{n_1}^{a_1}]J_{n_2}^{a_2}
+
J_{n_1}^{a_1}\mathcal L_\psi^\dagger[J_{n_2}^{a_2}]
-
\sum_{C,D,E=1}^N
\sum_{q\in\mathcal I}
\gamma(q,C)
t^{a_1}_{CD}t^{a_2}_{CE}
\psi^D_{n_1-q}\psi^E_{n_2+q}.
\end{aligned}
\label{eq:SM-two-current-eom-general}
\end{equation}
The first two terms are evaluated using Eq.~\eqref{eq:SM-single-current-general}, which consist of products of one current and one Majorana bilinear, yielding a maximal Majorana degree of four. The last term is inherently a Majorana bilinear with degree two. Thus, Eq.~\eqref{eq:SM-two-current-eom-general} exhibits the triangular structure $4\rightarrow 4,\;2,\;0 .$ Expanding the two currents gives
\begin{equation}
\begin{aligned}
O_2
=
\frac14
\sum_{A,B,C,D=1}^N
\sum_{p,q\in\mathcal I}
t^{a_1}_{AB}t^{a_2}_{CD}
:\psi^A_{n_1-p}\psi^B_p:
:\psi^C_{n_2-q}\psi^D_q: .
\end{aligned}
\label{eq:SM-O2-expanded}
\end{equation}
Using the canonical anticommutation relations (CARs), each product of normal-ordered bilinears is first reduced to a sum of ordered monomials of degrees four, two, and zero. The degree-two and degree-zero components are then solved using the bilinear equation and the identity $\mathcal L_\psi^\dagger[\mathbf 1]=0$, respectively.

We now present the finite-time evolution of a reduced ordered four-Majorana monomial. We consider $\psi^{A_1}_{r_1}\psi^{A_2}_{r_2}\psi^{A_3}_{r_3}\psi^{A_4}_{r_4}$, where repeated Majorana modes have already been reduced via the CARs. Defining the time-dependent contraction factor
\begin{equation}
\alpha_{A,r}(t)
=
\frac{\gamma(r,A)}{\Gamma_{r,A}}
\left(1-e^{-\Gamma_{r,A}t}\right),
\label{eq:SM-alpha-Ar-time}
\end{equation}
the Majorana triangular hierarchy and Eq.~\eqref{eq:SM-even-general-rule} dictate that
\begin{equation}
\begin{aligned}
\left[
\psi^{A_1}_{r_1}
\psi^{A_2}_{r_2}
\psi^{A_3}_{r_3}
\psi^{A_4}_{r_4}
\right](t)=&\;
\exp\left\{
\left[
-i\frac{2\pi v}{L}
(r_1+r_2+r_3+r_4)
-\frac12
\sum_{j=1}^4
\Gamma_{r_j,A_j}
\right]t
\right\}
\psi^{A_1}_{r_1}
\psi^{A_2}_{r_2}
\psi^{A_3}_{r_3}
\psi^{A_4}_{r_4}
\\
&+
\delta^{A_1A_2}\delta_{r_1+r_2,0}
\alpha_{A_1,r_1}(t)
\exp\left\{
\left[
-i\frac{2\pi v}{L}(r_3+r_4)
-\frac12
(\Gamma_{r_3,A_3}+\Gamma_{r_4,A_4})
\right]t
\right\}
\psi^{A_3}_{r_3}\psi^{A_4}_{r_4}
\\
&-
\delta^{A_1A_3}\delta_{r_1+r_3,0}
\alpha_{A_1,r_1}(t)
\exp\left\{
\left[
-i\frac{2\pi v}{L}(r_2+r_4)
-\frac12
(\Gamma_{r_2,A_2}+\Gamma_{r_4,A_4})
\right]t
\right\}
\psi^{A_2}_{r_2}\psi^{A_4}_{r_4}
\\
&+
\delta^{A_1A_4}\delta_{r_1+r_4,0}
\alpha_{A_1,r_1}(t)
\exp\left\{
\left[
-i\frac{2\pi v}{L}(r_2+r_3)
-\frac12
(\Gamma_{r_2,A_2}+\Gamma_{r_3,A_3})
\right]t
\right\}
\psi^{A_2}_{r_2}\psi^{A_3}_{r_3}
\\
&+
\delta^{A_2A_3}\delta_{r_2+r_3,0}
\alpha_{A_2,r_2}(t)
\exp\left\{
\left[
-i\frac{2\pi v}{L}(r_1+r_4)
-\frac12
(\Gamma_{r_1,A_1}+\Gamma_{r_4,A_4})
\right]t
\right\}
\psi^{A_1}_{r_1}\psi^{A_4}_{r_4}
\\
&-
\delta^{A_2A_4}\delta_{r_2+r_4,0}
\alpha_{A_2,r_2}(t)
\exp\left\{
\left[
-i\frac{2\pi v}{L}(r_1+r_3)
-\frac12
(\Gamma_{r_1,A_1}+\Gamma_{r_3,A_3})
\right]t
\right\}
\psi^{A_1}_{r_1}\psi^{A_3}_{r_3}
\\
&+
\delta^{A_3A_4}\delta_{r_3+r_4,0}
\alpha_{A_3,r_3}(t)
\exp\left\{
\left[
-i\frac{2\pi v}{L}(r_1+r_2)
-\frac12
(\Gamma_{r_1,A_1}+\Gamma_{r_2,A_2})
\right]t
\right\}
\psi^{A_1}_{r_1}\psi^{A_2}_{r_2}
\\
&+
\delta^{A_1A_2}\delta_{r_1+r_2,0}
\delta^{A_3A_4}\delta_{r_3+r_4,0}
\alpha_{A_1,r_1}(t)\alpha_{A_3,r_3}(t)\mathbf 1
\\
&-
\delta^{A_1A_3}\delta_{r_1+r_3,0}
\delta^{A_2A_4}\delta_{r_2+r_4,0}
\alpha_{A_1,r_1}(t)\alpha_{A_2,r_2}(t)\mathbf 1
\\
&+
\delta^{A_1A_4}\delta_{r_1+r_4,0}
\delta^{A_2A_3}\delta_{r_2+r_3,0}
\alpha_{A_1,r_1}(t)\alpha_{A_2,r_2}(t)\mathbf 1 .
\end{aligned}
\label{eq:SM-four-majorana-finite-time}
\end{equation}
The exact dynamics of the two-current operator is therefore constructed as follows. We first expand the operator
\begin{equation}
\begin{aligned}
O_2
=
\frac14
\sum_{A,B,C,D=1}^N
\sum_{p,q\in\mathcal I}
t^{a_1}_{AB}t^{a_2}_{CD}
:\psi^A_{n_1-p}\psi^B_p:
:\psi^C_{n_2-q}\psi^D_q: .
\end{aligned}
\label{eq:SM-O2-expanded-repeat}
\end{equation}
We then reduce each product of two normal-ordered bilinears into ordered Majorana monomials using the CARs. The degree-four monomials evolve according to Eq.~\eqref{eq:SM-four-majorana-finite-time}, the Majorana bilinears evolve according to Eq.~\eqref{eq:SM-bilinear-eq}, and the identity remains strictly constant. Thus, $O_2(t)$ is completely determined within the finite hierarchy $4,\;2,\;0$. Equivalently, combining Eq.~\eqref{eq:SM-O2-expanded} and Eq.~\eqref{eq:SM-four-majorana-finite-time} yields
\begin{equation}
\begin{aligned}
O_2(t)
=
\frac14
\sum_{A,B,C,D=1}^N
\sum_{p,q\in\mathcal I}
t^{a_1}_{AB}t^{a_2}_{CD}
\left[
:\psi^A_{n_1-p}\psi^B_p:
:\psi^C_{n_2-q}\psi^D_q:
\right](t),
\end{aligned}
\label{eq:SM-O2-finite-time-general}
\end{equation}
where the time evolution of each reduced monomial inside the square bracket is governed by Eq.~\eqref{eq:SM-four-majorana-finite-time} and the bilinear solution of Eq.~\eqref{eq:SM-bilinear-eq}.

Finally, we consider the long-time limit. Assuming that all relevant total decay rates are strictly positive,
\begin{equation}
\Gamma_{r,A}>0,\quad r>0,
\end{equation}
all homogeneous factors in Eq.~\eqref{eq:SM-four-majorana-finite-time} decay exponentially, leaving only the double-contraction terms. After carrying out the contractions across the current product, the long-time limit becomes
\begin{equation}
\begin{aligned}
\lim_{t\to\infty}
J_{n_1}^{a_1}J_{n_2}^{a_2}(t)
=&
\frac{\delta_{n_1+n_2,0}}{2}
\sum_{p\in\mathcal I}
\sum_{A,B=1}^N
t^{a_1}_{AB}t^{a_2}_{BA}\bar\alpha_{A,n_1-p}
\bar\alpha_{B,p}
\,\mathbf 1 ,
\end{aligned}
\label{eq:SM-O2-long-time-alpha}
\end{equation}
which depends exclusively on the stationary values of the Majorana oscillator contractions. Here, we defined
\begin{equation}
\bar\alpha_{A,r}
=
\lim_{t\to\infty}
\frac{\gamma(r,A)}{\Gamma_{r,A}}
\left(1-e^{-\Gamma_{r,A}t}\right)
=
\frac{\gamma(r,A)}{\Gamma_{r,A}} .
\end{equation}

If we apply the pure loss condition ($\gamma(r,A)>0$, $\gamma(-r,A)=0$ for $r>0$), $\bar\alpha_{A,r}$ reduces to the step function $\Theta(r)$, where \(\Theta(r)=1\) for \(r>0\) and \(\Theta(r)=0\) for \(r<0\). Consequently, for $n>0$, we obtain
\begin{equation}
\lim_{t\to\infty}
J_n^aJ_{-n}^b(t)
=
kn\delta^{ab}\mathbf 1,\quad\lim_{t\to\infty}
J_{-n}^aJ_n^b(t)
=
0.
\end{equation}
These results perfectly match the CFT vacuum expectation values of the affine currents. Because $J_n^a$ with $n>0$ annihilates the vacuum state, we have
\begin{equation}
\langle 0|J_n^aJ_{-n}^b|0\rangle
=
\langle 0|[J_n^a,J_{-n}^b]|0\rangle
=
kn\delta^{ab},\quad \langle 0|J_{-n}^aJ_n^b|0\rangle=0.
\end{equation}
The operator ordering governs the outcome: $J_n^aJ_{-n}^b$ captures the affine central contraction, whereas $J_{-n}^aJ_n^b$ is inherently normal-ordered with respect to the vacuum and thus decays entirely to zero.

\subsection{Ramond Sector}
\label{subsec:SM-Ramond-current}

In the Ramond sector, the presence of Majorana zero modes implies that the affine currents can contain zero-mode Majorana operators. We therefore decompose each current according to the number of zero modes it contains.

For the $n=0$ affine current, we write
\begin{equation}
J_0^a
=
J_{0,\mathrm{osc}}^a
+
J_{0,\mathrm{zero}}^a,
\end{equation}
where the oscillating- and zero-mode contributions are
\begin{equation}
\begin{aligned}
J_{0,\mathrm{osc}}^a
&=
\frac12
\sum_{A,B=1}^N
\sum_{\substack{q\in\mathbb Z\\ q\neq 0}}
t^a_{AB}
:\psi^A_{-q}\psi^B_q:,\quad J_{0,\mathrm{zero}}^a=
\frac12
\sum_{A,B=1}^N
t^a_{AB}
:\psi^A_0\psi^B_0: .
\end{aligned}
\label{eq:SM-Ramond-J0-split}
\end{equation}
For nonzero affine current modes ($n\neq 0$), we write
\begin{equation}
J_n^a
=
J_{n,\mathrm{osc}}^a
+
J_{n,\mathrm{mix}}^a,
\end{equation}
where the oscillating and mixed-mode components are given by
\begin{equation}
\begin{aligned}
J_{n,\mathrm{osc}}^a
&=
\frac12
\sum_{A,B=1}^N
\sum_{\substack{q\in\mathbb Z\\ q\neq 0,n}}
t^a_{AB}
:\psi^A_{n-q}\psi^B_q:,\quad J_{n,\mathrm{mix}}^a=
\frac12
\sum_{A,B=1}^N
t^a_{AB}
\left(
:\psi^A_n\psi^B_0:
+
:\psi^A_0\psi^B_n:
\right),
\end{aligned}
\label{eq:SM-Ramond-Jn-split}
\end{equation}
respectively. Using the antisymmetry of $t^a_{AB}$, the mixed part can be simplified to
\begin{equation}
J_{n,\mathrm{mix}}^a
=
\sum_{A,B=1}^N
t^a_{AB}\psi^A_n\psi^B_0,
\qquad n\neq 0.
\label{eq:SM-Ramond-mix-simple}
\end{equation}

In the oscillator part of $J_0^a$, each term consists of two nonzero Majorana modes. The identity source term in Eq.~\eqref{eq:SM-bilinear-eq} vanishes because it is proportional to $t^a_{AA}=0$. Thus, the time evolution is
\begin{equation}
\begin{aligned}
J_{0,\mathrm{osc}}^a(t)
=
\frac12
\sum_{A,B=1}^N
\sum_{\substack{q\in\mathbb Z\\ q\neq 0}}
t^a_{AB}
\exp\left[
-\frac12
\left(
\Gamma_{-q,A}
+
\Gamma_{q,B}
\right)t
\right]
:\psi^A_{-q}\psi^B_q: .
\end{aligned}
\label{eq:SM-Ramond-J0-osc-generic}
\end{equation}
For the zero-mode part, the corresponding jump operators act diagonally on reduced even zero-mode monomials. Therefore,
\begin{equation}
\begin{aligned}
J_{0,\mathrm{zero}}^a(t)
=
\frac12
\sum_{A,B=1}^N
t^a_{AB}
\exp\left[
-\left(
\gamma(0,A)+\gamma(0,B)
\right)t
\right]
:\psi^A_0\psi^B_0: .
\end{aligned}
\label{eq:SM-Ramond-J0-zero-generic}
\end{equation}
If zero-mode jumps $K_0^A$ are explicitly excluded, one simply sets $\gamma(0,A)=0$, ensuring that $J_{0,\mathrm{zero}}^a$ is strictly conserved. 

For $n\neq 0$, the oscillator component likewise consists of two nonzero Majorana modes, yielding the exact time evolution
\begin{equation}
\begin{aligned}
J_{n,\mathrm{osc}}^a(t)
=
\frac12
\sum_{A,B=1}^N
\sum_{\substack{q\in\mathbb Z\\ q\neq 0,n}}
t^a_{AB}
\exp\left[
\left(
-i\frac{2\pi v}{L}n
-\frac12\Gamma_{n-q,A}
-\frac12\Gamma_{q,B}
\right)t
\right]
:\psi^A_{n-q}\psi^B_q: .
\end{aligned}
\label{eq:SM-Ramond-Jn-osc-generic}
\end{equation}
The mixed part contains terms that couple one nonzero Majorana mode with one zero mode
\begin{equation}
\begin{aligned}
J_{n,\mathrm{mix}}^a(t)
=
\sum_{A,B=1}^N
t^a_{AB}
\exp\left[
\left(
-i\frac{2\pi v}{L}n
-\frac12\Gamma_{n,A}
-\gamma(0,B)
\right)t
\right]
\psi^A_n\psi^B_0 .
\end{aligned}
\label{eq:SM-Ramond-Jn-mix-generic}
\end{equation}
Again, if zero-mode jumps are absent, the final damping contribution is removed by setting $\gamma(0,B)=0$.

Equations~\eqref{eq:SM-Ramond-J0-osc-generic}-\eqref{eq:SM-Ramond-Jn-mix-generic} are exact for generic decay rates. They demonstrate that the Ramond current sector remains entirely governed by the Majorana bilinear hierarchy and closed within the conformal embedding subspace. Analogous to Eq.~\eqref{eq:SM-single-current-closed}, a uniform-rate condition explicitly recovers the closure of the current algebra for single-mode dynamics in the Ramond sector.

We now apply the uniform nonzero-mode limit, $\Gamma_{q,A}=\Gamma$ for $q\neq 0$, and omit zero-mode jumps by setting $\gamma(0,A)=0$. Under these conditions, the decomposed Ramond current sector closes into simplified forms
\begin{equation}
J_{0,\mathrm{osc}}^a(t)
=
e^{-\Gamma t}J_{0,\mathrm{osc}}^a,
\qquad
J_{0,\mathrm{zero}}^a(t)
=
J_{0,\mathrm{zero}}^a,
\end{equation}
and, for $n\neq 0$,
\begin{equation}
J_{n,\mathrm{osc}}^a(t)
=
e^{(-i2\pi vn/L-\Gamma)t}J_{n,\mathrm{osc}}^a,\quad J_{n,\mathrm{mix}}^a(t)
=
e^{(-i2\pi vn/L-\Gamma/2)t}J_{n,\mathrm{mix}}^a.
\end{equation}
Therefore, reconstructing the full currents gives
\begin{equation}
J_0^a(t)
=
e^{-\Gamma t}J_{0,\mathrm{osc}}^a
+
J_{0,\mathrm{zero}}^a,
\end{equation}
and, for $n\neq 0$,
\begin{equation}
\begin{aligned}
J_n^a(t)
=&
e^{(-i2\pi vn/L-\Gamma)t}J_{n,\mathrm{osc}}^a+
e^{(-i2\pi vn/L-\Gamma/2)t}J_{n,\mathrm{mix}}^a .
\end{aligned}
\end{equation}
Thus, under the condition of uniform nonzero-mode rates, the decomposed current sector closes explicitly. The resulting dynamics are exactly solvable, yielding respective decay rates of $\Gamma$, $\Gamma/2$, and $0$ for the oscillator, mixed, and zero-mode components.

\section{Verlinde-line Lindbladian and topological-charge dephasing}
\label{sec:SM-Verlinde-Lindbladian}

We now discuss another exactly solvable Lindbladian in CFT. This construction is independent of the Majorana-mode hierarchy derived in the previous two sections and applies to any rational conformal field theory (RCFT). This exact solvability arises because the jump operators are perfectly diagonal in the primary-sector (or topological-charge) basis.

\subsection{Primary-sector Subspace and TDL Jumps}
\label{subsec:SM-Verlinde-setup}
Consider a diagonal RCFT characterized by a finite set of primary labels $\mathcal P$. We restrict our focus to the primary-sector Hilbert space
\begin{equation}
\mathcal H_p
=
\mathrm{span}\{|a\rangle,\;a\in\mathcal P\}.
\label{eq:SM-primary-sector-Hilbert}
\end{equation}
The CFT Hamiltonian is diagonal in this basis:
\begin{equation}
H|a\rangle=\epsilon_a |a\rangle .
\label{eq:SM-primary-Hamiltonian}
\end{equation}
We define the primary state projectors and transition matrices as
\begin{equation}
P_a=|a\rangle\langle a|,
\qquad
P_{ab}=|a\rangle\langle b|,
\label{eq:SM-primary-projectors}
\end{equation}
which satisfy the matrix unit algebra
\begin{equation}
P_{ab}P_{cd}=\delta_{bc}P_{ad},
\qquad
P_aP_{bc}=\delta_{ab}P_{ac},
\qquad
P_{bc}P_a=\delta_{ca}P_{ba}.
\label{eq:SM-matrix-unit-algebra}
\end{equation}

Each primary label $x\in\mathcal P$ corresponds to a Verlinde topological defect line (TDL) $W_x$. These TDLs act diagonally on $\mathcal H_p$
\begin{equation}
W_x
=
\sum_{a\in\mathcal P}
\lambda_x(a)P_a,
\qquad
\lambda_x(a)
=
\frac{S_{xa}}{S_{\mathbf 1 a}}.
\label{eq:SM-Verlinde-eigenvalue}
\end{equation}
Here, $S_{xa}$ is the modular $S$-matrix, and $\mathbf 1$ denotes the vacuum primary. For a unitary diagonal RCFT, the Hermitian conjugate of a line is its charge-conjugate line
\begin{equation}
W_x^\dagger=W_{\bar x},
\qquad
\lambda_{\bar x}(a)=\lambda_x(a)^*.
\label{eq:SM-Verlinde-adjoint}
\end{equation}
We select these Verlinde lines as the Lindblad jump operators,
\begin{equation}
K_x=\sqrt{\eta_x}\,W_x,
\qquad
\eta_x\ge 0,
\label{eq:SM-Verlinde-jump}
\end{equation}
where the coefficient $\eta_x$ represents the measurement strength associated with the line $W_x$. The corresponding adjoint Lindbladian is
\begin{equation}
\mathcal L_V^\dagger[O]
=
i[H,O]
+
\sum_{x\in\mathcal P}
\eta_x
\left(
W_{\bar x}OW_x
-\frac12\{W_{\bar x}W_x,O\}
\right).
\label{eq:SM-Verlinde-adjoint-L}
\end{equation}
Because $H$, $W_x$, and $W_{\bar x}W_x$ are all simultaneously diagonalizable, the dynamics close exactly on the finite-dimensional operator space $\text{End}(\mathcal{H}_p)$ spanned by $\{P_{ab}\}$.

\subsection{Exact Solution}
\label{subsec:SM-Verlinde-exact-solution}
To fully determine the dynamics, it is sufficient to compute the action of $\mathcal L_V^\dagger$ on a single matrix unit $P_{ab}$, as any primary-sector operator can be expanded as
\begin{equation}
O
=
\sum_{a,b\in\mathcal P}
O_{ab}P_{ab}.
\label{eq:SM-primary-operator-expansion}
\end{equation}
The Hamiltonian contribution yields
\begin{equation}
i[H,P_{ab}]
=
i(\epsilon_a-\epsilon_b)P_{ab}.
\label{eq:SM-Verlinde-H-action}
\end{equation}
For the dissipative part, because $W_{\bar x}$ acts on the left index and $W_x$ acts on the right index, we have
\begin{equation}
W_{\bar x}P_{ab}W_x
=
\lambda_{\bar x}(a)\lambda_x(b)P_{ab}
=
\lambda_x(a)^*\lambda_x(b)P_{ab},
\label{eq:SM-Verlinde-sandwich}
\end{equation}
and
\begin{equation}
W_{\bar x}W_x
=
\sum_{c\in\mathcal P}
|\lambda_x(c)|^2P_c.
\label{eq:SM-Verlinde-WdaggerW}
\end{equation}
As a result,
\begin{equation}
W_{\bar x}W_xP_{ab}
=
|\lambda_x(a)|^2P_{ab},
\qquad
P_{ab}W_{\bar x}W_x
=
|\lambda_x(b)|^2P_{ab}.
\label{eq:SM-Verlinde-anticommutator-pieces}
\end{equation}
It follows that the anticommutator term evaluates to
\begin{equation}
\frac12\{W_{\bar x}W_x,P_{ab}\}
=
\frac12
\left(
|\lambda_x(a)|^2
+
|\lambda_x(b)|^2
\right)P_{ab}.
\label{eq:SM-Verlinde-anticommutator}
\end{equation}
Combining Eqs.~\eqref{eq:SM-Verlinde-H-action}, \eqref{eq:SM-Verlinde-sandwich}, and \eqref{eq:SM-Verlinde-anticommutator}, we obtain the full Lindbladian action
\begin{align}
\mathcal L_V^\dagger[P_{ab}]
=&\;
i(\epsilon_a-\epsilon_b)P_{ab}+
\sum_{x\in\mathcal P}
\eta_x
\left[
\lambda_x(a)^*\lambda_x(b)
-\frac12|\lambda_x(a)|^2
-\frac12|\lambda_x(b)|^2
\right]P_{ab}.
\label{eq:SM-Verlinde-Pab-action-raw}
\end{align}
We can now separate this complex coefficient into its real and imaginary components
\begin{equation}
\mathcal L_V^\dagger[P_{ab}]
=
\left(
-\Gamma_{ab}
+
i\omega_{ab}
\right)P_{ab},
\label{eq:SM-Verlinde-Pab-action}
\end{equation}
where the real decay rate is defined as
\begin{equation}
\Gamma_{ab}
=
\frac12
\sum_{x\in\mathcal P}
\eta_x
|\lambda_x(a)-\lambda_x(b)|^2,
\label{eq:SM-Verlinde-Gamma-ab}
\end{equation}
and the imaginary oscillation rate is
\begin{equation}
\omega_{ab}
=
\epsilon_a-\epsilon_b
+
\sum_{x\in\mathcal P}
\eta_x\,
\operatorname{Im}
\!\left[
\lambda_x(a)^*\lambda_x(b)
\right].
\label{eq:SM-Verlinde-omega-ab}
\end{equation}
Equation~\eqref{eq:SM-Verlinde-Gamma-ab} naturally follows from the algebraic identity
\begin{equation}
\operatorname{Re}
\left[
\lambda_x(a)^*\lambda_x(b)
-\frac12|\lambda_x(a)|^2
-\frac12|\lambda_x(b)|^2
\right]
=
-\frac12|\lambda_x(a)-\lambda_x(b)|^2.
\label{eq:SM-Verlinde-real-part-identity}
\end{equation}

Thus, each $P_{ab}$ is an eigenoperator of the adjoint Lindbladian. Its exact Heisenberg evolution is simply
\begin{equation}
P_{ab}(t)
=
\exp\!\left[
(-\Gamma_{ab}+i\omega_{ab})t
\right]P_{ab}.
\label{eq:SM-Verlinde-Pab-solution}
\end{equation}
For a general operator $O$ within the primary-sector space, Eq.~\eqref{eq:SM-primary-operator-expansion} gives
\begin{equation}
O(t)
=
\sum_{a,b\in\mathcal P}
O_{ab}
\exp\!\left[
(-\Gamma_{ab}+i\omega_{ab})t
\right]
P_{ab}.
\label{eq:SM-Verlinde-general-solution}
\end{equation}
For the diagonal projectors where $a=b$, we immediately find
\begin{equation}
\Gamma_{aa}=0,
\qquad
\omega_{aa}=0, \qquad \mathcal L_V^\dagger[P_a]=0.
\label{eq:SM-Verlinde-diagonal-zero}
\end{equation}
In summary, the Lindbladian strictly preserves the topological-charge probabilities while exponentially suppressing any off-diagonal coherences $P_{ab}$ ($a\neq b$) for which the corresponding decay rate $\Gamma_{ab}$ is strictly positive.

\subsection{Dephasing rate and distinguishability}
\label{subsec:SM-Verlinde-dephasing-rate}

Equation~\eqref{eq:SM-Verlinde-Gamma-ab} carries a direct physical measurement interpretation. The TDL $W_x$ can distinguish the primary sectors $a$ and $b$ if and only if
\begin{equation}
\lambda_x(a)\neq \lambda_x(b).
\end{equation}
If every active line in the environment satisfies
\begin{equation}
\lambda_x(a)=\lambda_x(b)
\qquad
\text{whenever }
\eta_x>0,\qquad\Gamma_{ab}=0,
\label{eq:SM-Verlinde-indistinguishable}
\end{equation}
the bath cannot distinguish the two topological-charge sectors, and the coherence between them is immune to damping, though it can still acquire a relative phase via $e^{i\omega_{ab}t}$. Conversely, if for every distinct pair $a\neq b$ there exists at least one active line $x$ such that
\begin{equation}
\eta_x>0,
\qquad
\lambda_x(a)\neq \lambda_x(b),
\label{eq:SM-Verlinde-distinguish-condition}
\end{equation}
then we strictly have
\begin{equation}
\Gamma_{ab}>0
\qquad
\text{for every }a\neq b.
\label{eq:SM-Verlinde-positive-Gamma}
\end{equation}
In this scenario, all off-diagonal primary-sector coherences decay exponentially, and Eq.~\eqref{eq:SM-Verlinde-general-solution} enforces the long-time limit
\begin{equation}
\lim_{t\to\infty}O(t)
=
\sum_{a\in\mathcal P}
O_{aa}P_a.
\label{eq:SM-Verlinde-long-time}
\end{equation}
This represents pure topological-charge dephasing: the environment continuously measures the topological charge while leaving its underlying probability distribution completely invariant.

The crucial distinction between invertible and noninvertible defect lines is also apparent in Eq.~\eqref{eq:SM-Verlinde-Gamma-ab}. If $x$ is an invertible line with unit quantum dimension $d_x=\lambda_{x}(\mathbf{1})=1$, the eigenvalues $\lambda_x(a)$ are necessarily U$(1)$ phases. Consequently, Eq.~\eqref{eq:SM-Verlinde-Gamma-ab} is strictly bounded from above, and the single-line decay contribution satisfies
\begin{equation}
\Gamma_{ab}^{(x)}
=
\frac12\eta_x
|\lambda_x(a)-\lambda_x(b)|^2\le 2\eta_x.
\label{eq:SM-Verlinde-single-line-rate}
\end{equation}
For a noninvertible line, however, the eigenvalues need not have a unit modulus, and its quantum dimension $\lambda_x(\mathbf 1)=d_x$ can exceed one. Thus, noninvertible lines are capable of producing significantly larger and more sector-selective dephasing rates, driven entirely by the eigenvalue separation $|\lambda_x(a)-\lambda_x(b)|$.

\subsection{Trace-distance dephasing bound}
\label{subsec:SM-Verlinde-trace-bound}

We now derive a trace-distance bound for this dephasing process. Let
\begin{equation}
\rho(t)=e^{t\mathcal L_V}\rho(0)
\end{equation}
denote the Schr\"odinger-picture density matrix. Its matrix elements in the primary-sector basis are
\begin{equation}
\rho_{ab}(t)
=
\langle a|\rho(t)|b\rangle
=
\operatorname{tr}\!\left(P_{ba}\rho(t)\right).
\label{eq:SM-Verlinde-rhoab-def}
\end{equation}
Exploiting the definition of the adjoint Lindbladian, we can write
\begin{equation}
\operatorname{tr}\!\left(P_{ba}\rho(t)\right)
=
\operatorname{tr}\!\left(P_{ba}(t)\rho(0)\right).
\label{eq:SM-Verlinde-duality}
\end{equation}
Substituting Eq.~\eqref{eq:SM-Verlinde-Pab-solution} gives
\begin{equation}
P_{ba}(t)
=
\exp\!\left[
(-\Gamma_{ba}+i\omega_{ba})t
\right]P_{ba}.
\end{equation}
Since the decay rates are symmetric ($\Gamma_{ba}=\Gamma_{ab}$), taking the absolute value yields
\begin{equation}
|\rho_{ab}(t)|
=
e^{-\Gamma_{ab}t}
|\rho_{ab}(0)|.
\label{eq:SM-Verlinde-rhoab-decay}
\end{equation}
This is the Schr\"odinger-picture decay law for the off-diagonal density matrix elements. The dynamical phase is governed by $\omega_{ba}$, but it drops out completely upon taking the absolute value.

We define the diagonal projection of a density matrix as
\begin{equation}
\Pi(\rho)
=
\sum_{a\in\mathcal P}
P_a\rho P_a.
\label{eq:SM-Verlinde-diagonal-projection}
\end{equation}
Because $P_a(t)=P_a$, this diagonal part is strictly conserved
\begin{equation}
\Pi(\rho(t))=\Pi(\rho(0)).
\label{eq:SM-Verlinde-diagonal-conserved}
\end{equation}
The remaining off-diagonal component is
\begin{equation}
X(t)
=
\rho(t)-\Pi(\rho(0))
=
\sum_{\substack{a,b\in\mathcal P\\a\neq b}}
\rho_{ab}(t)P_{ab}.
\label{eq:SM-Verlinde-offdiagonal-X}
\end{equation}
Assuming the active lines successfully distinguish all primary sectors, we identify the slowest off-diagonal dephasing rate
\begin{equation}
\Gamma_V
=
\min_{a\neq b}\Gamma_{ab}.
\label{eq:SM-Verlinde-GammaV}
\end{equation}
Equation~\eqref{eq:SM-Verlinde-rhoab-decay} then imposes the strict inequality
\begin{equation}
|\rho_{ab}(t)|
\le
e^{-\Gamma_V t}
|\rho_{ab}(0)|,
\qquad
a\neq b.
\label{eq:SM-Verlinde-rhoab-bound}
\end{equation}
Therefore, the Hilbert-Schmidt norm of $X(t)$ is bounded by
\begin{align}
\|X(t)\|_2^2
&=
\sum_{a\neq b}
|\rho_{ab}(t)|^2\le
e^{-2\Gamma_V t}
\sum_{a\neq b}
|\rho_{ab}(0)|^2
=
e^{-2\Gamma_V t}
\|X(0)\|_2^2.
\label{eq:SM-Verlinde-HS-bound}
\end{align}
For any properly normalized density matrix, the Hilbert-Schmidt norm satisfies
\begin{equation}
\|\rho(0)\|_2^2
=
\operatorname{tr}\rho(0)^2
\le 1.
\end{equation}
Since $X(0)$ is merely the off-diagonal projection of $\rho(0)$, it inherits the bound
\begin{equation}
\|X(0)\|_2\le 1.
\end{equation}
Thus, we arrive at
\begin{equation}
\|X(t)\|_2
\le
e^{-\Gamma_V t}.
\label{eq:SM-Verlinde-HS-final}
\end{equation}
The trace distance to the diagonal projection is defined as
\begin{equation}
D\!\left(\rho(t),\Pi(\rho(0))\right)
=
\frac12
\left\|
\rho(t)-\Pi(\rho(0))
\right\|_1
=
\frac12
\|X(t)\|_1.
\label{eq:SM-Verlinde-trace-distance-def}
\end{equation}
For an operator acting on the \(|\mathcal P|\)-dimensional Hilbert space \(\mathcal H_p\), standard norm inequalities give
\begin{equation}
\|X(t)\|_1
\le
\sqrt{|\mathcal P|}\,\|X(t)\|_2.
\label{eq:SM-Verlinde-trace-HS}
\end{equation}
Combining Eqs.~\eqref{eq:SM-Verlinde-HS-final}, \eqref{eq:SM-Verlinde-trace-distance-def}, and \eqref{eq:SM-Verlinde-trace-HS}, we obtain the final trace-distance bound
\begin{equation}
D\!\left(\rho(t),\Pi(\rho(0))\right)
\le
\frac{\sqrt{|\mathcal P|}}{2}
e^{-\Gamma_V t}.
\label{eq:SM-Verlinde-trace-bound}
\end{equation}
Let $0<\delta<1$ represent a prescribed trace-distance accuracy tolerance. We define the dephasing time $t_{\rm deph}(\delta)$ as the precise time after which
\begin{equation}
D\!\left(\rho(t),\Pi(\rho(0))\right)\le \delta.
\end{equation}
Equation~\eqref{eq:SM-Verlinde-trace-bound} supplies the sufficient condition
\begin{equation}
\frac{\sqrt{|\mathcal P|}}{2}
e^{-\Gamma_V t}
\le
\delta.
\end{equation}
Solving this for $t$ yields a rigorous upper bound for the required dephasing time
\begin{equation}
t_{\rm deph}(\delta)
\le
\max\left\{
0,\,
\frac{1}{\Gamma_V}
\log
\frac{\sqrt{|\mathcal P|}}{2\delta}
\right\}.
\label{eq:SM-Verlinde-dephasing-time}
\end{equation}
The maximum with zero trivially accounts for scenarios where the requested accuracy is already looser than the bound at $t=0$. The critical physical takeaway is that the dephasing time is entirely dictated by the inverse of the slowest distinguishable dephasing rate, $\Gamma_V$, and grows only logarithmically with the desired precision $1/\delta$.

\subsection{Example: Ising CFT}
\label{subsec:SM-Verlinde-Ising}

We conclude with the simplest example: the Ising CFT, whose primary labels are
\begin{equation}
\mathcal P=\{\mathbf 1,\sigma,\psi\}.
\end{equation}
In the ordered basis $(\mathbf 1,\sigma,\psi)$, the modular $S$-matrix is
\begin{equation}
S
=
\frac12
\begin{pmatrix}
1 & \sqrt2 & 1\\
\sqrt2 & 0 & -\sqrt2\\
1 & -\sqrt2 & 1
\end{pmatrix}.
\label{eq:SM-Ising-S-matrix}
\end{equation}
Using the relation
\begin{equation}
\lambda_x(a)
=
\frac{S_{xa}}{S_{\mathbf 1 a}},
\end{equation}
we obtain the following Verlinde-line eigenvalues:
\begin{center}
\begin{tabular}{c|ccc}
\hline\hline
line \(x\)
&
\(\lambda_x(\mathbf 1)\)
&
\(\lambda_x(\sigma)\)
&
\(\lambda_x(\psi)\)
\\
\hline
\(\mathbf 1\)
&
\(1\)
&
\(1\)
&
\(1\)
\\
\(\psi\)
&
\(1\)
&
\(-1\)
&
\(1\)
\\
\(\sigma\)
&
\(\sqrt2\)
&
\(0\)
&
\(-\sqrt2\)
\\
\hline\hline
\end{tabular}
\end{center}
The identity line yields the same eigenvalue across every sector and therefore causes no dephasing.

For each line $x\in \mathcal P$, the single-line decay contribution is given by
\begin{equation}
\Gamma_{ab}^{(x)}
=
\frac12\eta_x
|\lambda_x(a)-\lambda_x(b)|^2.
\label{eq:SM-Ising-single-line-rate}
\end{equation}
The three independent pairs of primary sectors are
\begin{equation}
(\mathbf 1,\sigma),
\qquad
(\mathbf 1,\psi),
\qquad
(\sigma,\psi).
\end{equation}
Their corresponding single-line dephasing rates evaluate to
\begin{center}
\begin{tabular}{c|ccc}
\hline\hline
line \(x\)
&
\(\Gamma_{\mathbf 1,\sigma}^{(x)}\)
&
\(\Gamma_{\mathbf 1,\psi}^{(x)}\)
&
\(\Gamma_{\sigma,\psi}^{(x)}\)
\\
\hline
\(\mathbf 1\)
&
\(0\)
&
\(0\)
&
\(0\)
\\
\(\psi\)
&
\(2\eta_\psi\)
&
\(0\)
&
\(2\eta_\psi\)
\\
\(\sigma\)
&
\(\eta_\sigma\)
&
\(4\eta_\sigma\)
&
\(\eta_\sigma\)
\\
\hline\hline
\end{tabular}
\end{center}
Thus, if both the $\psi$ and $\sigma$ lines are active in the environment, the total dephasing rates are
\begin{equation}
\Gamma_{\mathbf 1,\sigma}
=
\eta_\sigma+2\eta_\psi,
\label{eq:SM-Ising-Gamma-1sigma}
\end{equation}
\begin{equation}
\Gamma_{\mathbf 1,\psi}
=
4\eta_\sigma,
\label{eq:SM-Ising-Gamma-1psi}
\end{equation}
and
\begin{equation}
\Gamma_{\sigma,\psi}
=
\eta_\sigma+2\eta_\psi.
\label{eq:SM-Ising-Gamma-sigmapsi}
\end{equation}
Therefore, the slowest dephasing rate governing the system is
\begin{equation}
\Gamma_V^{\rm Ising}
=
\min\{
\eta_\sigma+2\eta_\psi,\,
4\eta_\sigma
\}.
\label{eq:SM-Ising-GammaV}
\end{equation}
As $|\mathcal P|=3$, Eq.~\eqref{eq:SM-Verlinde-dephasing-time} yields the dephasing time bound
\begin{equation}
t_{\rm deph}^{\rm Ising}(\delta)
\le
\max\left\{
0,\,
\frac{1}{\Gamma_V^{\rm Ising}}
\log
\frac{\sqrt3}{2\delta}
\right\},
\label{eq:SM-Ising-deph-time}
\end{equation}
provided that $\Gamma_V^{\rm Ising}>0$. Equivalently, the dephasing time for the Ising CFT can be written explicitly as
\begin{equation}
t_{\rm deph}^{\rm Ising}(\delta)
\le\max\left\{
0,\,
\frac{1}{\min\{\eta_\sigma+2\eta_\psi,\;4\eta_\sigma\}}
\log
\frac{\sqrt3}{2\delta}
\right\}.
\label{eq:SM-Ising-deph-time-explicit}
\end{equation}
This example clearly illustrates the physical distinguishability criterion. If $\eta_\sigma>0$, the noninvertible $\sigma$ line distinguishes all three primary sectors, driving complete primary-sector dephasing. Conversely, if only the invertible $\psi$ line is active, then
\begin{equation}
\Gamma_{\mathbf 1,\psi}=0,
\end{equation}
meaning the coherence between the $\mathbf 1$ and $\psi$ sectors does not dephase. In this scenario, the Lindbladian achieves only partial dephasing of the Ising primary-sector Hilbert space.

\section{Virasoro \(\text{SL}_q(2,\mathbb{R})\) Lindbladian dynamics}
\label{sec:SM-Virasoro-SLq}

We now present an explicit and mathematically rich example of exact operator closure within a finite subalgebra of the infinite-dimensional Virasoro algebra. By carefully selecting the jump operators, we can restrict the dynamics to a closed operator sector generated by the finite set of Virasoro modes $\{L_0, L_q, L_{-q}\}$, where $q$ is a uniquely fixed positive integer.

Throughout this derivation, we use $\text{SL}_q(2,\mathbb{R})$ as a convenient shorthand for this $q$-rescaled Virasoro $\mathfrak{sl}(2,\mathbb{R})$ block. It is crucial to emphasize that this notation strictly denotes a classically rescaled subalgebra of the standard Virasoro generators, not a quantum-group $q$-deformation. We restrict our analysis to a single chiral sector to maintain clarity, the anti-chiral (anti-holomorphic) sector is completely symmetric and can be obtained via the straightforward replacement $L_n\rightarrow \bar L_n$.

The infinite-dimensional Virasoro algebra is given by
\begin{equation}
[L_m,L_n]
=
(m-n)L_{m+n}
+
\frac{c}{12}m(m^2-1)\delta_{m+n,0}\mathbf 1 .
\label{eq:SM-Vir-algebra}
\end{equation}
By restricting our focus to a fixed excitation mode $q>0$, Eq.~\eqref{eq:SM-Vir-algebra} perfectly isolates a finite triad of commutation relations
\begin{equation}
[L_0,L_q]=-qL_q,
\quad
[L_0,L_{-q}]=qL_{-q},\quad [L_q,L_{-q}]
=
2qL_0+C_q\mathbf 1,
\quad
C_q
=
\frac{c}{12}q(q^2-1).
\label{eq:SM-Vir-L0-Lpm}
\end{equation}
Here, $C_q$ represents the structurally critical central extension contribution evaluated at the mode $q$. Equivalently, by absorbing this c-number central term into a redefinition of the primary scaling operator, we find that the operators
\begin{equation}
L_q,\quad L_{-q},\quad
\widetilde L_0
:=
L_0+\frac{c}{24}(q^2-1)\mathbf 1,
\label{eq:SM-Vir-shifted-L0}
\end{equation}
which form a \(q\)-rescaled \(\mathfrak{sl}(2,\mathbb{R})\) algebra.
Indeed, one can verify that the shifted generator explicitly satisfies
\begin{equation}
[L_q,L_{-q}]=2q\widetilde L_0,
\qquad
[\widetilde L_0,L_q]=-qL_q,
\qquad
[\widetilde L_0,L_{-q}]=qL_{-q}.
\label{eq:SM-Vir-slq}
\end{equation}
This is the fundamental underlying mechanism that permits a closed Virasoro block to manifest under open Lindbladian dynamics without triggering an infinite operator cascade.

We choose the chiral Hamiltonian to be
\begin{equation}
H_{\text{ch.}}=\Omega L_0,
\qquad
\Omega>0,
\label{eq:SM-Vir-H}
\end{equation}
where \(\Omega\) is the level spacing per unit \(L_0\).
The two Lindblad jump operators are linear in non-zero Virasoro modes
\begin{equation}
K_q=\sqrt{\gamma_+}\,L_q,
\qquad
K_{-q}=\sqrt{\gamma_-}\,L_{-q},
\qquad
\gamma_\pm\geq 0.
\label{eq:SM-Vir-jumps}
\end{equation}
Physically, $K_q$ and $K_{-q}$ represent environmental channels that absorb and inject conformal excitations of momentum $q$, respectively, with independent coupling rates $\gamma_+$ and $\gamma_-$. The exact adjoint Lindbladian governing the Heisenberg-picture dynamics of an arbitrary chiral observable $O$ is therefore given by
\begin{equation}
\begin{split}
\mathcal L_q^\dagger[O]
&=
i[H_{\text{ch.}},O]
+
\gamma_+
\left(
L_{-q}OL_q
-\frac12\{L_{-q}L_q,O\}
\right)+
\gamma_-
\left(
L_qOL_{-q}
-\frac12\{L_qL_{-q},O\}
\right).
\end{split}
\label{eq:SM-Vir-adjoint-Lindbladian}
\end{equation}
We also introduce the deformed Virasoro Hamiltonian inside this fixed \(q\)-block as the observable,
\begin{equation}
G_q(\alpha)
=
\alpha_0L_0
+
\alpha_+\frac{L_q+L_{-q}}{2}
+
\alpha_-\frac{L_q-L_{-q}}{2i},
\label{eq:SM-Vir-Gq}
\end{equation}
where \(\alpha_0,\alpha_+\), and \(\alpha_-\) are real.

For later use, we first introduce a useful operator identity.
For a jump operator \(A\), the Heisenberg dissipator can be written as
\begin{equation}
A^\dagger O A-\frac12\{A^\dagger A,O\}
=
\frac12 A^\dagger[O,A]
+
\frac12[A^\dagger,O]A.
\label{eq:SM-Vir-diss-comm-form}
\end{equation}
Deploying Eq.~\eqref{eq:SM-Vir-diss-comm-form} alongside the $\text{SL}_q(2,\mathbb{R})$ commutation relations, the action of the Lindbladian on the fundamental generators $L_0, L_q, L_{-q}$ is evaluated as follows
\begin{equation}
\mathcal L_q^\dagger[L_0]
=
q\gamma_-L_qL_{-q}
-
q\gamma_+L_{-q}L_q,
\label{eq:SM-Vir-generic-L0}
\end{equation}
\begin{equation}
\begin{split}
\mathcal L_q^\dagger[L_q]
&=
-i\Omega qL_q
+
\frac{\gamma_-}{2}
L_q(2qL_0+C_q\mathbf 1)-
\frac{\gamma_+}{2}
(2qL_0+C_q\mathbf 1)L_q,
\end{split}
\label{eq:SM-Vir-generic-Lq}
\end{equation}
and
\begin{equation}
\begin{split}
\mathcal L_q^\dagger[L_{-q}]
&=
i\Omega qL_{-q}
+
\frac{\gamma_-}{2}
(2qL_0+C_q\mathbf 1)L_{-q}-
\frac{\gamma_+}{2}
L_{-q}(2qL_0+C_q\mathbf 1).
\end{split}
\label{eq:SM-Vir-generic-Lminusq}
\end{equation}
Equations~\eqref{eq:SM-Vir-generic-L0}--\eqref{eq:SM-Vir-generic-Lminusq} demonstrate that, for generic asymmetric rates where $\gamma_+ \neq \gamma_-$, the operator space spanned by $\{\mathbf 1, L_0, L_q, L_{-q}\}$ is not closed. Under time evolution, nonlinear operator products such as $L_{-q}L_q$, $L_qL_{-q}$, and $L_0L_{\pm q}$ are generated. Thus, the generic-rate dynamics is hard to track analytically.

The finite block closes when the two rates are symmetric,
\begin{equation}
\gamma_+=\gamma_-=\gamma .
\label{eq:SM-Vir-symmetric-rates}
\end{equation}
In this case, Eq.~\eqref{eq:SM-Vir-generic-L0} gives
\begin{equation}
\begin{split}
\mathcal L_q^\dagger[L_0]
&=
\gamma q(L_qL_{-q}-L_{-q}L_q)=
\gamma q[L_q,L_{-q}]=
2\gamma q^2L_0
+
\frac{\gamma c}{12}q^2(q^2-1)\mathbf 1.
\end{split}
\label{eq:SM-Vir-sym-L0}
\end{equation}
Similarly, Eqs.~\eqref{eq:SM-Vir-generic-Lq} and
\eqref{eq:SM-Vir-generic-Lminusq} reduce to
\begin{equation}
\mathcal L_q^\dagger[L_q]
=
(\gamma q^2-i\Omega q)L_q,\quad \mathcal L_q^\dagger[L_{-q}]
=
(\gamma q^2+i\Omega q)L_{-q}.
\label{eq:SM-Vir-sym-Lq}
\end{equation}
Therefore, the dynamics of $G_q(\alpha)$ is closed within $\operatorname{span}\{\mathbf 1,L_0,L_q,L_{-q}\}$ and is exactly solvable. The shifted generator \(\widetilde L_0\) defined in
Eq.~\eqref{eq:SM-Vir-shifted-L0} obeys the homogeneous equation
\begin{equation}
\mathcal L_q^\dagger[\widetilde L_0]
=
2\gamma q^2\widetilde L_0.
\label{eq:SM-Vir-shifted-L0-dynamics}
\end{equation}
Consequently, the exact Heisenberg solutions are
\begin{equation}
L_0(t)
=
e^{2\gamma q^2t}L_0(0)
+
\frac{c}{24}(q^2-1)
\left(e^{2\gamma q^2t}-1\right)\mathbf 1,
\label{eq:SM-Vir-L0-solution}
\end{equation}
\begin{equation}
L_q(t)
=
e^{(\gamma q^2-i\Omega q)t}L_q(0),\quad L_{-q}(t)
=
e^{(\gamma q^2+i\Omega q)t}L_{-q}(0).
\label{eq:SM-Vir-Lq-solution}
\end{equation}
Combining Eqs.~\eqref{eq:SM-Vir-L0-solution}--\eqref{eq:SM-Vir-Lq-solution}, we obtain the exact solution for the deformed Hamiltonian:
\begin{equation}
\begin{split}
G_q(\alpha;t)
&=
\alpha_0
\left[
e^{2\gamma q^2t}L_0(0)
+
\frac{c}{24}(q^2-1)
\left(e^{2\gamma q^2t}-1\right)\mathbf 1
\right]
\\
&\quad
+
\frac{\alpha_+}{2}
\left[
e^{(\gamma q^2-i\Omega q)t}L_q(0)
+
e^{(\gamma q^2+i\Omega q)t}L_{-q}(0)
\right]
\\
&\quad
+
\frac{\alpha_-}{2i}
\left[
e^{(\gamma q^2-i\Omega q)t}L_q(0)
-
e^{(\gamma q^2+i\Omega q)t}L_{-q}(0)
\right].
\end{split}
\label{eq:SM-Vir-Gq-solution}
\end{equation}
In the large-time limit $t\gg 1$, as $\gamma>0$ and $q>0$, $G_q(\alpha;t)$ is dominated by the $\widetilde L_0$ term, which grows exponentially as $e^{2\gamma q^2t}$.

Finally, let us contrast the symmetric closed channel with a single-jump cooling channel.
If one takes
\begin{equation}
\gamma_+=\gamma,
\qquad
\gamma_-=0,
\label{eq:SM-Vir-single-jump}
\end{equation}
then Eq.~\eqref{eq:SM-Vir-generic-L0} gives
\begin{equation}
\mathcal L_{q,+}^\dagger[L_0]
=
-\gamma qL_{-q}L_q.
\label{eq:SM-Vir-single-jump-L0}
\end{equation}
Since \(L_{-q}=L_q^\dagger\) in a unitary CFT, the expectation value of \(L_0\) decreases under this single \(L_q\) jump.
This is the zero-temperature, \(q\)-mode cooling direction.
However, the same single-jump choice does not preserve the finite
\(\operatorname{span}\{\mathbf 1,L_0,L_q,L_{-q}\}\) block.
For example, Eq.~\eqref{eq:SM-Vir-generic-Lq} becomes
\begin{equation}
\mathcal L_{q,+}^\dagger[L_q]
=
-i\Omega qL_q
-
\frac{\gamma}{2}
(2qL_0+C_q\mathbf 1)L_q,
\label{eq:SM-Vir-single-jump-Lq}
\end{equation}
which contains a product of Virasoro generators.
Thus cooling and finite \(\text{SL}_q(2,\mathbb{R})\) closure are separated in this construction: the symmetric channel is exactly closed but heating, while the single-jump channel is cooling-like but not closed in the finite Virasoro block.

In this section, we restrict our observables to linear combinations of Virasoro modes within a single fixed-$q$ subalgebra. However, this framework can be naturally extended to include broader classes of observables, such as the modes of primary operators $O_{h}$ with conformal dimension $h$. These extensions will be explored in detail in an upcoming work by Wen et al. \cite{UWW}.

\end{widetext}

\end{document}